\documentclass[journal=jctcce,manuscript=article,layout=traditional]{achemso} 
\usepackage[T1]{fontenc} 
\usepackage[version=3]{mhchem} 
\usepackage[dvipsnames]{xcolor}
\usepackage{graphicx,subcaption}
\usepackage{amssymb}
\usepackage{overpic}
\usepackage{braket}
\usepackage{comment}
\usepackage{soul}
\usepackage{bm}
\usepackage{hyperref}
\hypersetup{pdfborderstyle={/S/U/W 0.5}}
\usepackage{booktabs}
\usepackage{enumitem}
\usepackage{multirow}
\usepackage{tikz}
\usepackage{svg}
\usepackage{siunitx}

\usepackage{bibunits}
\defaultbibliographystyle{unsrt}

\newcommand{\hH}{{\hat{H}}}
\newcommand{\hJ}{{\hat{J}}}
\newcommand{\hrho}{{\hat{\rho}}}
\newcommand{\pp}[2]{\frac{\partial{#1}}{\partial{#2}}}
\newcommand{\hX}{{\hat{X}}}
\newcommand{\hY}{{\hat{Y}}}
\newcommand{\hP}{{\hat{P}}}

\newcommand{\hO}{{\hat{O}}}
\newcommand{\hbm}[1]{\hat{\bm{#1}}} 
\newcommand{\hU}{{\hat{U}}}
\newcommand{\hL}{{\hat{L}}}

\newcommand{\bF}{{\bf F}}
\newcommand{\bP}{{\bf P}}

\newcommand{\bd}{{\bf d}}

\newcommand{\bnabla}{{\bm \nabla}}

\newcommand{\hbP}{{\hat{\bf P}}}

\newcommand{\hbp}{{\hat{\bf p}}}
\newcommand{\hbd}{{\hat{\bf d}}}
\newcommand{\hbr}{{\hat{\bf r}}}
\newcommand{\hbR}{{\hat{\bf R}}}

\newcommand{\hLambda}{{\hat{\Lambda}}}
\newcommand{\hbGamma}{{\hat{\bf \Gamma}}}
\newcommand{\tbGamma}{{\tilde{\bf \Gamma}}}

\newcommand{\bS}{{\bf S}}

\newcommand{\bR}{{\bf R}}
\newcommand{\br}{{\bf r}}
\newcommand{\bV}{{\bf V}}

\newcommand{\bp}{{\bf p}}

\newcommand{\joeb}[1]{\textcolor{black}{#1}}

\author{Joseph E. Subotnik}
\email{subotnik@princeton.edu}
\affiliation{Department of Chemistry, Princeton University, Princeton, New Jersey 08544}
\author{Ethan Alguire}
\affiliation{Schrodinger Inc., NY, NY 10036}
\author{Nicole Bellonzi}
\affiliation{Apollo Quantum LLC, Cambridge, MA 02142}
\author{Xuezhi Bian}
\affiliation{Department of Chemistry, Princeton University, Princeton, New Jersey 08544}
\author{Mansi Bhati}
\affiliation{Department of Chemistry, Princeton University, Princeton, New Jersey 08544}
\author{Nadine Bradbury}
\affiliation{Department of Chemistry, Princeton University, Princeton, New Jersey 08544}
\author{D. Vale Cofer-Shabica}
\affiliation{Department of Chemistry, Princeton University, Princeton, New Jersey 08544}
\author{Ben Curlee}
\affiliation{Department of Chemistry, Princeton University, Princeton, New Jersey 08544}
\author{Titouan Duston}
\affiliation{Department of Chemistry, Princeton University, Princeton, New Jersey 08544}
\author{Shervin Fatehi}
\affiliation{Slater Matsil LLP, Dallas, Texas 75252}
\author{Gaohan Miao}
\affiliation{}
 
\author{Zheng Pei}
\affiliation{Department of Chemistry, 
Brandeis University,
Waltham, MA 02453-2728}

\author{Linqing Peng}
\affiliation{Department of Chemistry, Princeton University, Princeton, New Jersey 08544}
\author{Tian Qiu}
\affiliation{Department of Chemistry, Princeton University, Princeton, New Jersey 08544}
\author{Michael Rosen}
\affiliation{Department of Chemistry, Princeton University, Princeton, New Jersey 08544}
\author{Zhen Tao}
\affiliation{Department of Chemistry, 
University of Rhode Island, Kingston, RI 02881}
\author{Hung-Hsuan Teh}
\affiliation{The Institute for Solid State Physics, The University of Tokyo, Kashiwa,
Chiba, 277-8581, Japan}
\affiliation{Graduate School of Informatics, Nagoya University, Nagoya, 464-0814, Japan}
\author{Xinchun Wu}
\affiliation{Department of Chemistry, Princeton University, Princeton, New Jersey 08544}
\author{Yanze Wu}
\affiliation{Department of Chemistry, Northwestern University, 
Evanston, IL 60208}
\author{Zain Zaidi}
\affiliation{Department of Chemistry, Princeton University, Princeton, New Jersey 08544}

\author{Yihan Shao}
\affiliation{Department of Chemistry, 
Brandeis University,
Waltham, MA 02453-2728}

\author{Jonathan Rawlinson}
\affiliation{Department of Mathematics, Nottingham Trent University,
Nottingham
NG1 4FQ, UK}

\author{Neil Shenvi}
\affiliation{The
South Durham School of Science and Math, Durham, NC 27713}

\author{Robert Littlejohn}
\affiliation{Department of Physics, University of California, Berkeley, CA, 94720}

\title{A Perspective on Phase Space Electronic Structure Theory -- From Its  Surface Hopping Origins Through To Its Future Promise }

\begin{document}

\maketitle
\pagebreak

\begin{abstract}
    We trace the history of phase space electronic structure theory (PSEST),  highlighting how this powerful approach emerged from fundamental questions in semiclassical surface hopping dynamics and evolved into an alternative to standard Born-Oppenheimer based electronic structure theory (with moving instead of frozen nuclei). Our goal herein is not to recapitulate the mathematical details of phase space electronic structure calculations, and few equations are presented so as to maximize readability. Instead, our goal is to provide intuition for those new to the field -- both  (i) regarding the physics present when solving the Schrodinger equation in a non-inertial frame as well as (ii) regarding why PSEST is a necessary step forward towards understanding chemical problems involving spin (with limited practical alternatives).   We further highlight some of the many open questions in this fast developing area, which will hopefully inspire new practitioners in this field.  This intuitive perspective lacks many equations and is meant to complement (rather than replace) the more technical review given in Bian {\em et al}, {\em Chem. Phys. Rev.}  {\bf 7}, 011303 (2026).
\end{abstract}

\section{Introduction}

\begin{figure}[h]
    \centering
    \includegraphics[width=1.0\linewidth]{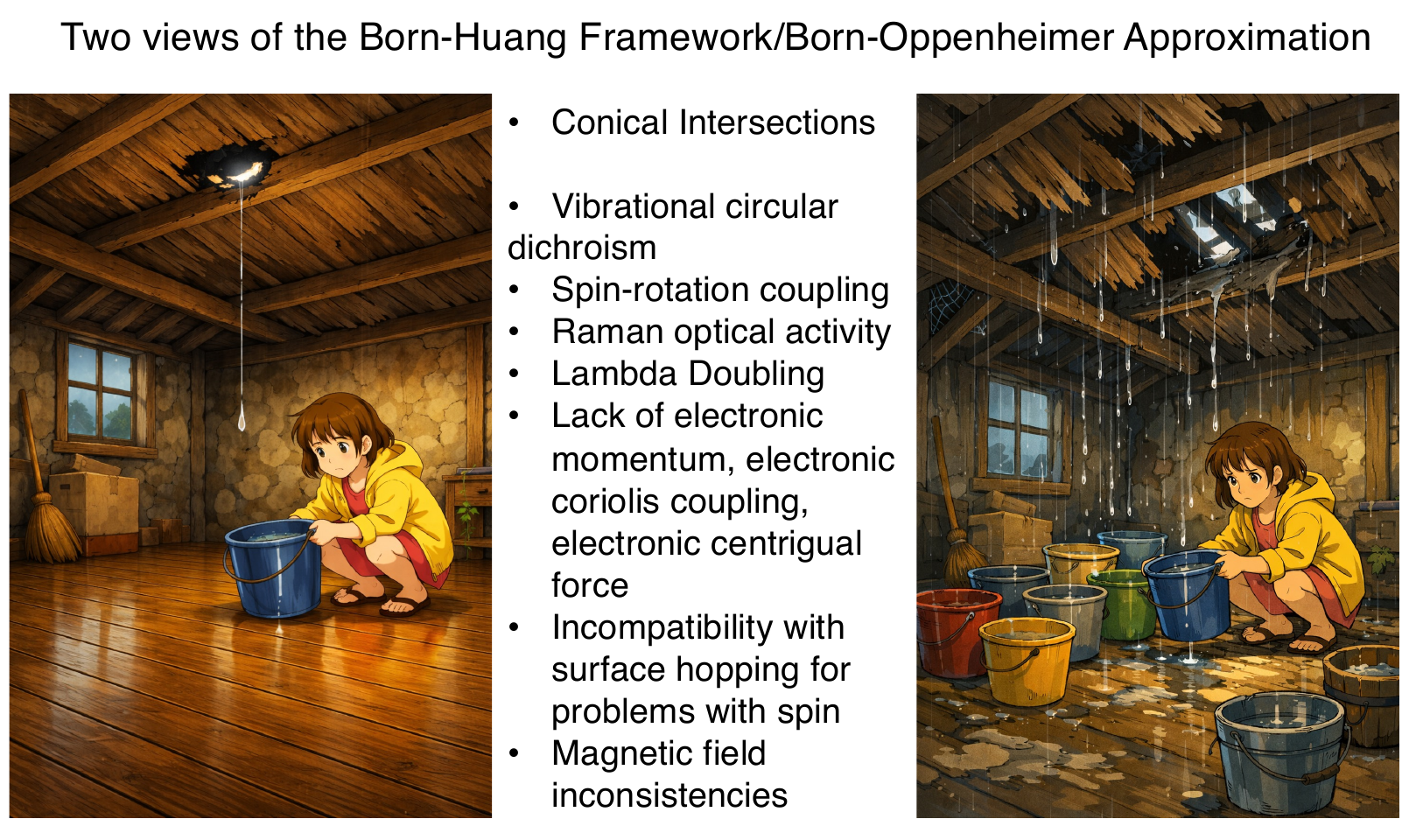}
    \caption{{\em (Left)}: For many problems in chemistry and physics, the Born-Huang framework   offers an efficient means to extract any and all observable quantities; in fact, the only important failure for running dynamics within the BO approximation is the presence of conical intersections.   Within such a metaphor, the Born Oppenheimer approximation would appear to be a strong house with just one hole in its roof.    {\em (Right)}: For many other problems in chemistry and physics, because the Born-Oppenheimer approximation has so many smaller failures (as listed above in the middle panel), a better analogy would be  a shack with a very leaky roof.  Credit: ChatGPT where we asked AI to create an image of a child in a room with a roof with either  one hole or a roof with many holes. }
    \label{fig:linqing}  
\end{figure}

Over the last two decades, phase space electronic structure theory (PSEST) has emerged as a major focus of the research developments of many of the present authors.  From our perspective, this transformation has 
been a slow evolution, rather than a rapid change of state.  Nevertheless, 
for the most part, the chemical physics community is still adjusting to a non-Born Oppenheimer view of not just chemical dynamics but also electronic structure. The goal of this account is to offer physical chemists an intuitive, but also personal, introduction to phase space electronic structure theory. Our intention is to guide the reader historically through some of the discoveries and realizations  that  forced us to develop such a non-Born Oppenheimer based approach to the electronic structure theory problem (see Fig. \ref{fig:linqing}), a very non-conventional field of study at this moment.

There is always the question of where to begin a historical narrative --- nothing comes from nothing. Our perspective is  that, in our hands, the direct driver for phase space electronic structure theory came from questions that were not answerable within standard surface hopping calculations, and so our narrative will begin in 1990, with Tully's seminal surface hopping paper\cite{tully:fssh,tully:faraday:fssh}.  Because our goal here is to be as intuitive as possible, we will aim to include as few equations as necessary and introduce the relevant concepts slowly (without expecting much background).

The authors are well aware that Born-Oppenheimer theory\cite{born_oppenheimer_1927} is the cornerstone of modern chemistry, i.e.  the theory upon which all of our intuition in chemistry is based\cite{jensen}, and we also can appreciate  that wrapping one's head around phase space electronic structure theory can be painful. The hope of the authors is that, as future generations mature and learn physical chemistry, some of the ideas presented here will be less non-intuitive and perhaps, even, mainstream.
Note that this historical perspective should not be considered as an alternative to the more comprehensive and technical review in Ref. \citenum{xuezhi:cpr:review:2026}.  Whereas Ref. \citenum{xuezhi:cpr:review:2026} summarizes what exactly phase space electronic structure theory is, the present perspective is intended to offer more context, especially historical context, explaining where PSEST comes from and why we believe the method is necessary for practical problems going forward.

\subsection{The Born-Huang Framework, the Born-Oppenheimer Approximation, and Surface Hopping}\label{sec:tully}

\subsubsection{The Born-Huang Framework vs. the Born-Oppenheimer Approximation}

\begin{figure}
    \centering
    \includegraphics[width=\linewidth]{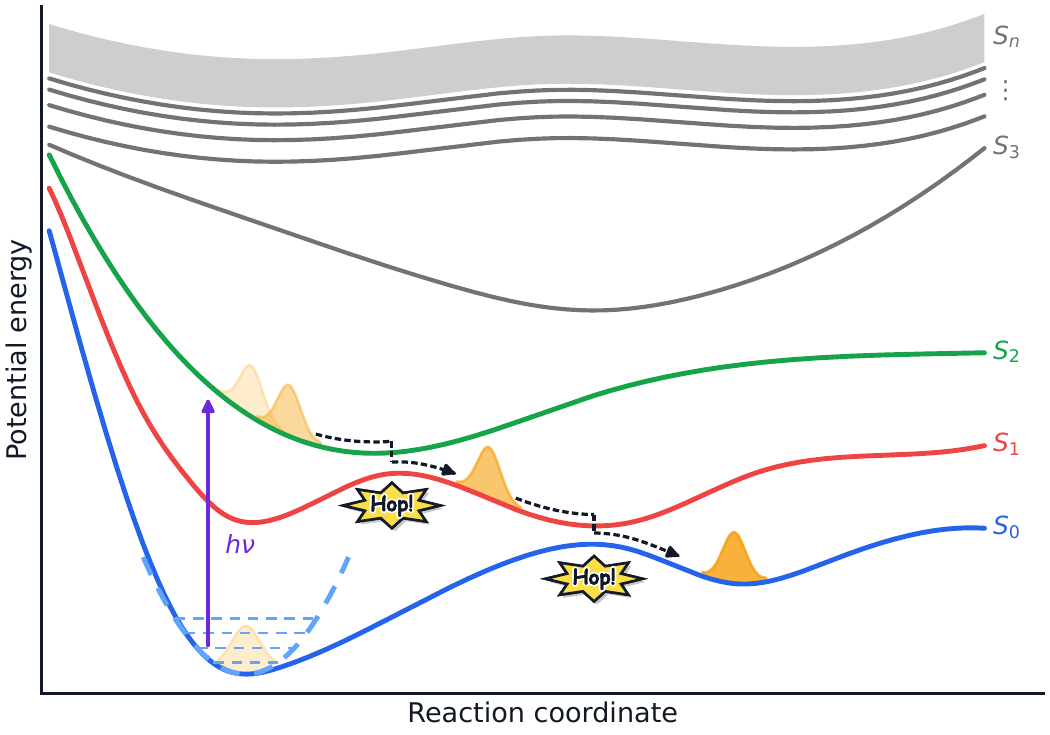}
    \caption{A typical scenario for photochemistry: a molecule is photo-excited to the second excited state ($S_2$), relaxes downhill, jumps to the first excited state ($S_1$), relaxes further, and then crosses back to the ground state ($S_0$) in a different chemical configuration.  Note that here we label electronic states adiabatically, i.e. $S_2(\bR)$ is always the third lowest energy  state. Most surface hopping algorithms will run on only a few states (here, it would seem four states are natural, $S_0,S_1,S_2,$ and $S_3$.)}
    \label{fig:pes}
\end{figure}

All quantum chemistry textbooks\cite{szabo:ostlund} begin with the Born-Huang framework\cite{born:huang} and the Born-Oppenheimer approximation\cite{born_oppenheimer_1927}.  The Born-Huang {\em framework} specifies that, in order to understand chemistry,  we freeze the heavy nuclei and we solve the Schr\"{o}dinger equation for the stationary states of the electronic degrees of freedom.  Mathematically, if the nuclear positions are denoted $\bR$ and the momenta are denoted $\bP$, we write $\hH_{tot} = \hbP^2/(2M) + \hH_{el}(\bR)$  and we diagonalize $\hH_{el}(\bR)$, 
\begin{eqnarray}
    \hH_{el}(\bR) \Phi_n(\br;\bR) = E_n(\bR) \Phi_n(\br;\bR)
    \label{eq:diagHBO}
\end{eqnarray}(Freezing the nuclei means setting the nuclear kinetic energy to zero, $\hbP^2/(2M) = 0$.)
The result of the above diagonalization is a set of potential energy surfaces $E_n(\bR)$ as shown in Figure \ref{fig:pes}.
Here, $E_0(\bR)$ is the ground state potential energy surface,  $E_1(\bR)$ is the first excited state potential energy surface, etc.
These potential energy surfaces are also called {\em adiabatic} surfaces in the literature.
The Born-Oppenheimer {\em approximation} then dictates that nuclei move alone one such level (usually $E_0(\bR)$) at a time. When one models the vibration of a diatomic molecule as a harmonic 
oscillator, one is implicitly using the Born-Oppenheimer approximation; if we were to rely on the heuristic model in Fig. \ref{fig:pes}, 
the harmonic oscillator's force constant comes from the shape of $E_0(\bR)$ at the global minimum.

Now, note that,  as shown in the figure,  very often curves tend to cross. For instance, note the crossing   between the first and second excited states.  These crossings can either be \joeb{locally} avoided or conical and have been studied for  many decades\cite{truhlar:mead:2003:ci_vs_avoided}; they are very common indeed and raise a critical question: how do molecules move on surfaces when those surfaces appear to coincide and/or cross? 
This basic question has inspired chemical physicists for decades (and continues to do). The phenomena of particles  moving along multiple potential energy surfaces  is  referred to as "nonadiabatic dynamics" within chemical physics, and this phenomena underlies the nonadiabatic limit of Marcus's theory of  electron transfer\cite{marcus:1956} (which holds in the limits that the donor and acceptor are well separated\cite{nitzanbook}).
As we will discuss below, understanding such phenomena was a prime motivator  for the development of phase space electronic structure theory (PSEST) in our laboratories. 

Is there an intuitive, heuristic approach to
solving the dynamics that appear at a curve crossing, one that can be explained to a general chemistry reader who does not want to look at equations or rigorously solve for the coupled
nuclear-electronic wavefunction that emerges from a degenerate or nearly degenerate point in space? Yes.  
 For most physical chemists, especially experimentalists, a surface hopping view 
 (credited to John Tully\cite{tully:2000:review}) usually suffices.  As an example of how surface hopping works, consider a photo excited molecule  moving along surface 2  in Fig. \ref{fig:pes}. When the molecule comes to the crossing, the particle can either stay on surface 2 or jump to surface 1 and continue relaxing downward.  Eventually, after the trajectory reaches another crossing point, the particle hops down again, and returns to the ground state surface, $E_0(\bR)$ (and in this case, a chemical reaction has occurred).  This cartoon version of  radiationless electronic relaxation  is commonly accepted nowadays as correct to zeroth order (and there are several means to make it rigorous\cite{subotnik:2013:qcle_fssh_derive} including the capacity to launch full spawning calculations\cite{martinez:1996:jcp_nai}).

In a moment, we will  review the most salient points of the surface hopping approach.  Before beginning such a review, however, it is critical to remind the reader that, even though the surface hopping algorithm just described represents a non-adiabatic approach to chemical dynamics that goes beyond the Born-Oppenheimer {\em approximation},  the entire surface hopping algorithm {\em is} based on the Born-Huang {\em framework}:  the adiabatic  potential energy surfaces in Fig. \ref{fig:pes}  are generated by solving Eq. \ref{eq:diagHBO}.

\subsubsection{Tully's Surface Hopping}

Newtonian dynamics along a given potential energy surface have been well understood for hundreds of years. Thus, in developing his surface hopping approach, 
Tully's major accomplishment was to predict a general rate of hopping from  one surface to another (in a fashion that was compatible with Newtonian dynamics on one surface before and after the hop).
Following Ref. \citenum{tully:fssh},
in order to derive Tully's proposed rate of hopping, imagine one wants to follow the 
dynamics of a trajectory initialized on one adiabatic PES $j$ (say, $j =  2$ in Fig. \ref{fig:pes}). Suppose further that one is running calculations for the first four electronic states ($S_0$, $S_1$, $S_2$ and $S_3$ again as in Fig. \ref{fig:pes}).  Intuitively, one can write down the instantaneous electronic wave function at any time $t$ as
\begin{eqnarray}
\label{eq:psi_expand}
    \Psi(\br,t) = \sum_{n = 0}^3 c_n(t) \Phi_n(\br;\bR(t))
\end{eqnarray}
Crucially, note that the adiabatic eigenstates $\left\{ \Phi_n \right\}$ depend on nuclear position $\bR$, 
$\Phi_n(\br;\bR(t))$. 

In order to calculate the amplitudes $c_k$ for every $k$, one asserts the electronic Schr\"{o}dinger equation $i \hbar d\ket{\Psi}/dt = \hH_{el} \ket{\Psi}$ and  plugs in Eq. \ref{eq:psi_expand} above. The result is:

\begin{eqnarray}
& &    i\hbar \dot{\Psi}(\br,t) = i\hbar \left(\sum_{n = 0}^3 \dot{c}_n(t) \Phi_n(\br;\bR(t)) + \sum_{n = 0}^3 c_n(t) \dot{\Phi}_n(\br;\bR(t))\right) = \hH_{el} \sum_{n = 0}^3 c_n(t) \Phi_n(\br;\bR(t))
\nonumber \\
\\
    \implies & & i\hbar \sum_{n = 0}^3 \dot{c}_n(t) \Phi_n(\br;\bR(t))  = \sum_{n = 0}^3 c_n(t) \hH_{el}  \Phi_n(\br;\bR(t)) -  i\hbar \sum_{n = 0}^3 c_n(t) \dot{\bR}\frac{\partial \Phi_n}{\partial \bR}
\end{eqnarray}

If we now multiply by $\Phi_k^*$ and integrate over $\br$, and define the nuclear momentum as $\bP = M\dot{\bR}$
the result is:

\begin{eqnarray}
i\hbar \dot{c}_k(t) =  \sum_n \left(
 H^{el}_{kn}(\bR)  c_n -
i\hbar \frac{{\bP}}{M} \cdot \bd_{kn}(\bR)  c_n \right)
\label{eq:etdse}
\end{eqnarray}
Of course, this result can be reduced further because, in the adiabatic basis, $H_{kn} = E_k \delta_{kn}$ (see Eq. \ref{eq:diagHBO}), but let's use the notation in Eq. \ref{eq:etdse} for now for reasons that will become clear later.
In Eq. \ref{eq:etdse}, we have defined the derivative coupling as 
\begin{eqnarray}
    \bd_{kn} = \left< \Phi_k \middle| \frac{\partial \Phi_n}{\partial \bR} \right>
\end{eqnarray}

According to Eq. \ref{eq:etdse} above, transitions from state $k$ to state $n$ are determined by the derivative coupling $\bd_{kn}$.   In plain spoken English, 
the rate of electronic relaxation and hopping depends on how quickly the electronic character of states $k$ and $n$ mix and so,  in his seminal work\cite{tully:fssh}, Tully predicted a hopping rate from $k \rightarrow n$ that was proportional to $\bd_{kn} \cdot \bP$. Finally, to fix the constant of proportionality, Tully insisted that the hopping rate should be as minimal as possible (i.e. there should be as few hops as possible)  in order to ensure that the  number of trajectories on surface $k$  ($N_k$) divided by the number of trajectories on surface $n$ ($N_n$) be equal to the ratio of the electronic populations:
\begin{eqnarray}
    \frac{N_k}{N_n} = \frac{|c_k|^2}{|c_n|^2}
    \label{eq:consistency}
    \end{eqnarray}
  The result of this analysis is a hopping rate of the form $\gamma_{k \rightarrow n} = 2 \mbox{Re} \left( c_n c_k^* \bd_{kn} \cdot \bP/ M \right)/|c_k|^2 $  and the resulting fewest switches surface hopping (FSSH) algorithm agrees with experiment quite often\cite{tretiak:2011:conj,landry:2014:abinitio_closs,barbatti:2011:review}. 

 There is now one more important fact about surface hopping that must be mentioned. Namely, surface hopping conserves energy: when a system relaxes from an excited state (e.g. state 2) to a lower electronic state (e.g., state 1), the resulting  loss of   potential energy must be compensated by a gain in the kinetic energy  of the nuclei; in other words, if the electrons relax, the nuclei heat up. But which nuclei should heat up?  Based on arguments of Pechukas\cite{pechukas:1969:pr} and Herman\cite{herman:1984:jcp:rescaling_direction},  Tully hypothesized that the change in momentum should be in the direction of the derivative coupling. Mathematically, Tully conjectured that 
 \begin{eqnarray}
\bP_{new} = \bP_{old} +\alpha {\bf d}     
\label{eq:rescale}
 \end{eqnarray}
where $\alpha$ is determined by energy conservation.  This ansatz was later confirmed by Kapral\cite{kapral:1999:jcp} through comparison with higher-level theories\cite{subotnik:2016:arpc,kapral:2016:chemphys_fssh}.

There are many more nuances associated with surface hopping and many reviews have been written on this topic\cite{hynes:ci_fssh:review,subotnik:2016:arpc}, but the picture above should be sufficient for the time being.  Again, the algorithm just described has been largely successful in terms of understanding many photochemical experiments\cite{barbatti:2011:review}, reflecting the notion that Tully's search for a minimal hopping rate was physically meaningful.

\subsection{Shenvi's Superadiabatic Approach}\label{sec:shenvi}

Following up on Tully's 1990 paper\cite{tully:fssh}, in 2009  Shenvi hypothesized \cite{shenvi:2009:jcp_pssh}
 that one could construct potential energy surfaces for non-adiabatic problems with even smaller hopping rates than what Tully had been able to achieve.  Considering Eq. \ref{eq:etdse}, Shenvi's  
 idea was that, rather than work with adiabatic surfaces from diagonalizing $\hH_{el}$ in Eq. \ref{eq:diagHBO},  why not diagonalize (or  rather rediagonalize) the matrix
 in Eq. \ref{eq:Hshenvi}, 
 \begin{eqnarray}
 \label{eq:Hshenvi}
     \hH_{Shenvi} = \frac{(\bP - i\hbar \hbd)^2}{2M} + \hH^{el}_{diag} 
 \end{eqnarray}
and then hop between the resulting eigenstates?
After all, Eq. \ref{eq:Hshenvi}  includes the $\bP \cdot \bd$ terms that drive electronic transitions; so why not construct electronic states that already take those transitions into account?
 
Inspired by  previous work by Berry on a slightly different problem involving adiabatic motion on periodic time dependent potentials\cite{berry:1987:superadiabat}, Shenvi
called the resulting eigenstates of this operator superadiabats; today, we might also call them phase space adiabats or phase space  potential energy surfaces. Below, we will simply refer to these surfaces as Shenvi's phase space surfaces, 
\begin{eqnarray}
\label{eq:diagHShenvi}
\hH_{Shenvi}(\bR,\bP) \Psi_n^{Shenvi}(\br;\bR,\bP) = E_n(\bR,\bP) \Psi_n^{Shenvi}(\br;\bR,\bP)
\end{eqnarray}

As noted above, Shenvi  was motivated  by the goal of minimizing the hopping rate beyond what Tully had been able to achieve; the price Shenvi had to pay was that
his potential energy surfaces now depended on both $\bR$ and $\bP$; if $N$ is the number of nuclei, these PESs  occupy a $6N$ not $3N$ dimensional parameter space. 
That being said, in Ref. \citenum{shenvi:2009:jcp_pssh}, Shenvi was able to prove that this ansatz could recover certain features  that were simply impossible using Born Oppenheimer surfaces.  One of Shenvi’s favorite hamiltonians was the following
\begin{eqnarray}
    \hat{H}_{flat} = 
    \frac{\hP_X^2}{2M} + 
    \left(
\begin{array}{cc}
\cos(\theta(\hX)) & \sin(\theta(\hX)) \\
\sin(\theta(\hX)) & -\cos(\theta(\hX))
\end{array}
    \right)
    \label{eq:Hflat}
\end{eqnarray}
Here,  the adiabatic potential energy surfaces are always flat (for all $X$) because the eigenvalues of the electronic hamiltonian in Eq. \ref{eq:Hflat} are $\pm 1$,  independent of the form of $\theta$. Thus, $\hat{H}_{flat}$ is a very approachable model problem where one can easily isolate dynamical effects.

Unfortunately, although Shenvi's phase space surface hopping (PSSH) showed strong potential in many ways (see especially the work of Izmaylov\cite{izmaylov:2016:jpc_dboc_pssh}), to our knowledge, no one in the chemistry community ever ran {\em ab initio} electronic structure calculations (or much less ran {\em ab initio} phase-space surface hopping dynamical simulations).   In practice, the procedure outlined by Shenvi (with the diagonalization in Eq. \ref{eq:diagHShenvi}) is simply too expensive.  To run Shenvi's PSSH dynamics, first one must  generate a set of BO surfaces and then second  differentiate those wavefunctions to calculate derivative couplings and then third, if one wants to run dynamics, one must diagonalize again and then further differentiate the resulting surfaces. The Shenvi scheme is impractical.   Furthermore, Ref. \citenum{izmaylov:2016:jpc_dboc_pssh} found  that for some model problems, phase space surface hopping (PSSH) 
performed worse than Tully's FSSH based on BO theory.   At root, the problem is that near a conical intersection,  where the derivative coupling $\bd$ becomes very large, a phase space potential surface will not look at all like the BO potential surface and  can have huge barriers and/or sinks which are often unphysical. In this very nonadiabatic limit, where one must hop between surfaces, Shenvi's phase space approach would appear to be a step in the wrong direction.
Finally, perhaps the most important omission  of Shenvi’s earlier work  is that he never thoroughly studied multi-dimensional problems, but rather focused mostly on one-dimensional systems.  As a result, the operator $\bd$ was largely one dimensional and  could not fully represent a magnetic field,  a theme to which we will return.  

 Before concluding this section,  it is worth noting one more problem with Shenvi's  PSSH approach, namely the question of degeneracy. 
 For the case that two electronic states are meaningfully degenerate (say, two Kramers doublets),   Shenvi's PSSH approach is doomed to fail  because electronic adiabats are not well defined in that case.   Thus, if one were to attempt to implement PSSH,  one would necessarily generate an arbitrary  choice of adiabats, with an arbitrary $\bd$ matrix, and the resulting dynamics would be similarly arbitrary.  By contrast, Tully's FSSH algorithm would still be stable (because the adiabatic surfaces are still well-defined).

\subsection{Wigner/Weyl transform}

Before we can go any further and discuss the jump from Shenvi's original PSSH approach to a more modern view of phase space electronic structure theory, we must introduce one concept that is often not well known  to young chemical physicists who are not familiar with Miller's semiclassical hierarchy\cite{miller:2001:jpcareview}, namely the concept of a Wigner-Weyl transform\cite{case:2008:wigner_review}. As background, note that after much debate, one of the standard rules for theoretical chemistry is  (unfortunately) that there is no unique semi-classical means to  embed classical mechanics within quantum dynamics\cite{miller:2012:coherence}; classical and quantum variables do not play naturally with each other. That being said, the Wigner-Weyl transforms do offer a  well-defined path for taking the classical limit of a quantum problem.  Although Shenvi himself did not invoke Wigner-Weyl transforms in Ref. \citenum{shenvi:2009:jcp_pssh}, a decade earlier, Ciccotti and Kapral had indeed constructed an early surface hopping model of chemical dynamics based on Wigner-Weyl transforms\cite{kapral:1999:jcp}.

 Wigner-Weyl transforms  work as follows. Given a quantum mechanical operator $\hO$  -- for example, which could be $\hbR$ or $\hbP$ -- one can construct a ``symbol'' in phase space through a Wigner transform as follows.
\begin{eqnarray} 
    \hat O_W(\bR,\bP) &=& 
    \int d\bm R' \bra{\bm R + \frac {\bm R'} 2 } \hat O \ket{\bm R - \frac {\bm R'} 2 } e^{-\frac i \hbar \bm R' \cdot \bm P} \label{eq:wigner} 
\end{eqnarray}
Crucially, one must recognize that this transformation is invertible through a Weyl transform, which can be constructed as follows:
\begin{eqnarray} 
    \bra{\bm R} \hat{O} \ket{\bm R'}  
    & = & \int \frac {d\bm P} {2\pi\hbar} e^{\frac i\hbar \bm P \cdot (\bm R - \bm R')} \hat O_W\left(\frac {\bm R + \bm R'} 2, \bm P\right)
        \label{eq:weyl}
\end{eqnarray}
Wigner-Weyl transforms are analogues of fourier transforms for operators.

At the end of the day, through the use of Wigner-Weyl transforms,  one has the power to accomplish several useful tasks.  On the one hand, one can use a Wigner transform to visualize and/or conceptualize quantum operators in phase space. For instance, the density matrix corresponding to a complex gaussian wavefunction becomes a real-valued gaussian wavepacket in phase space\cite{tannor:quantumbook}. 
On the other hand and more importantly,  if one starts with a quantum wave function for a system with electrons and nuclei, one can then construct the partial Wigner transform (transforming only over the nuclei) and build a well-defined (though again, not unique) model for semi-classical dynamics where the nuclei are treated classically and the electrons are treated quantum mechanically\cite{coker:2012:iterative,kapral:2008:jcp_pbme}.  Indeed, Kapral and Cicotti based their own semiclassical surface hopping approach\cite{kapral:1999:jcp} on such a    transformation. Indeed, we will also follow this route below.

Before completing this brief introduction to Wigner-Weyl transforms, for what follows below, we must emphasize that the Wigner transform does not commute with multiplication\cite{tannor:quantumbook}. In other words, for two operators $\hat{A}$ and $\hat{B}$,  $(\hat{A} \hat{B})_W \ne \hat{A}_W \hat{B}_W$. Instead, there is a much more complicated formula:
\begin{eqnarray}
(\hat{A} \hat{B})_W =  \hat{A}_W *\hat{B}_W.
\label{eq:Wstar}
\end{eqnarray}
Because we will not do much math below, we will not bother to define  the star $(*)$ product here---it involves the Poisson bracket and the exponential operator and can rarely be evaluated exactly in practice. But it is important that the star product is associative:
$(\hat{A}_W  * \hat{B}_W) * \hat{C}_W
 = \hat{A}_W  *(\hat{B}_W  * \hat{C}_W)
 =
 \hat{A}_W *\hat{B}_W * \hat{C}_W$.

This section completes our review of the necessary background material for PSEST.

\section{The Momentum Problem in Standard Surface Hopping}

\subsection{An Old Problem: Momentum Conservation and ETFs}\label{sec:hop_etf}

As one can tell from the description in Sec. \ref{sec:tully}, the  central element of nonadiabatic dynamics, and specifically of the surface hopping algorithm,   is the derivative coupling $\bd$.   Now, within the realm of multireference electronic structure theory focused on curve crossings,  the derivative couplings between many body electronic states were calculated in the 1980s, as pioneered by Lengsfield  and Yarkony and others\cite{yarkony:1984:jcp_dercouple,yarkony:1992:advchemphys}.   Interestingly, however, in 2011,   during a fairly straightforward calculation of the analogous derivative of couplings between simple configuration interaction single (CIS) states,  our research group stumbled upon  on a problem with surface hopping's approach to derivative couplings\cite{fatehi:2011:dercouple}  -- a problem centered around the question of momentum that had long been partially neglected.

The problem originates when one constructs an {\em ab initio} calculation for the derivative couplings in an atomic orbital basis.  Therein, one finds that for each and every calculation, the final result for $\bd_{jk}$ can always be written down as the sum of two different terms:
\begin{align}
\label{eq:etf0}
    \bd_{jk} &= \bd_{jk}^0 + \bd_{jk}^{ETF} \\
    \bd_{jk}^{I, ETF} &=  \sum_{\mu \nu p q} \tilde{\bS}^I_{\mu \nu} C_{\mu p} C_{\nu q} \left< \Phi_j \middle | a_p^{\dagger} a_q \middle | \Phi_k\right> 
    \label{eq:etf} 
\end{align}
Here, we have used standard  quantum chemistry notation to label $\mu$ and $\nu$ as atomic orbitals, whereas $p$ and $q$ index molecular orbitals; $\left< \Phi_j \middle | a_p^{\dagger} a_q \middle | \Phi_k\right>$ is the one-electron transition density matrix between the CIS states $\Phi_j$ and $\Phi_k$; and $C_{\mu p}$ is the matrix of coefficients expressing molecular orbitals in terms of atomic orbitals. Finally, $\tilde{\bS}$ is the antisymmetrized overlap matrix, 
\begin{eqnarray}
 \tilde{\bS}^I_{\mu \nu} =  \frac{1}{2} \left( \left< \mu \middle| \partial \nu/\partial \bR_I \right> -
\left< \partial \mu/\partial \bR_I \middle| \nu \right> \right) 
\end{eqnarray}
Importantly, if one sums over all of the atoms in the molecular system, one finds that the first term vanishes $\sum_I \bd^{I,0}_{jk} = 0$;  however, if one performs the same summation for the second term, the result is not zero but rather exactly proportional to the electronic momentum:
\begin{eqnarray}
    \sum_I 
    \bd^I_{jk} =     \sum_I 
    \bd^{I,ETF}_{jk} = 
    \sum_I \sum_{\mu \nu} \tilde{\bS}^I_{\mu \nu} C_{\mu p} C_{\nu q} \left< \Phi_j \middle | a_p^{\dagger} a_q \middle | \Phi_k\right> = \frac{1}{i \hbar} \bp^e_{jk} \ne 0 
    \label{eq:early_etf}
\end{eqnarray}
When combined with the surface hopping ansatz for momentum scaling in Eq. \ref{eq:rescale}, Eq. \ref{eq:early_etf}  would seem to predict some very unphysical scenarios. For instance,  and most outrageously,
consider the case of a hydrogen atom that is excited to the $2p_x$ state moving in the x-direction. For such a case, there is a nonzero derivative coupling, $d_{1s,2p_x}$, so that there is the possibility that the electron will jump down from the $2p_x$ state to the $1s$ state and start translating faster in the $x-$direction. Clearly, this scenario is incorrect: an excited hydrogen atom can travel without relaxing. More broadly, the algorithm is  confused about what to do with the electronic momentum: the algorithm is not able to conserve both energy and momentum. 


In Refs. \citenum{fatehi:2011:dercouple,fatehi:2012:dercouple}, we developed an intuitive model to explain the problem. Let the many-body states of interest be $\ket{\Phi_i}$ and $\ket{\Phi_j}$ which have well-defined density matrices and suppose (as usual) that the  molecular orbitals  are linear combinations of atomic orbitals:
\begin{eqnarray}
    \phi_i(\br) = \sum_{\mu_A} \chi_{\mu_A}(\br) C_{\mu_A i}
\end{eqnarray}
Here, we have explicitly made clear that each atomic orbital $\chi_\mu$ is associated with one nucleus, here denoted $A$. 
In order to model the fact the nuclei are moving, one can boost every atomic orbital in the wavefunction with the velocity of the corresponding nucleus, $\bV_A$, and replace  $\phi_i(\br)$ with $\tilde{\phi}_i(\br)$, where
\begin{eqnarray}
    \tilde{\phi}_i(\br) = \sum_{\mu_A} \chi_{\mu_A}(\br) e^{im_e \bV_A \cdot \br/\hbar} C_{\mu_A i} \approx
    \sum_{\mu_A}
    \left(1 + \frac{im_e \bV_A \cdot \br}{\hbar}\right)\chi_{\mu_A}(\br) 
 C_{\mu_A i} 
\end{eqnarray}
At this point, one must be prepared to make approximations and, for example, ignore the fact that $\tilde{\phi}_i$ and $\tilde{\phi}_j$ are not strictly orthogonal; in particular, if we let $\tilde{\Phi}_i$ and $\tilde{\Phi}_j$ be the many-body states with boosted atomic orbitals, let us assume that these states are also orthonormal. 
If one now substitutes $\tilde{\Phi}_k$ for $\Phi_k$ when building up the matrix element of $\hH_{el}$ in Eq. \ref{eq:etdse}, a new term arises 
(because $\ket{\tilde{\phi}_k}$  is not an eigenvector of the fock matrix -- $\ket{\phi_k}$ is the eigenvector). 
Namely, one can show that 
\begin{eqnarray}
\tilde{H}_{ij} \equiv \left<\tilde{\Phi}_i \middle | \hH \middle | \tilde{\Phi}_j\right> \approx 
\left<\Phi_i \middle | \hH \middle | \Phi_j\right> + i  \hbar \sum_{A \mu \nu p q} \bV_A \tilde{\bS}^A_{\mu \nu} C_{\mu p} C_{\nu q} \left< \Phi_i \middle | a_p^{\dagger} a_q \middle | \Phi_j\right> 
\label{eq:newH}
\end{eqnarray}

Finally, note that if one plugs  Eq. \ref{eq:newH}  and \ref{eq:etf} into Eq. \ref{eq:etdse} (replacing $H_{kn}$ with $H_{\tilde{k} \tilde{n}}$), the two extraneous terms cancel and we find that we must replace $\bd$ with $\bd^0$:
\begin{eqnarray}
i\hbar \dot{c}_k(t) =  \sum_n \left(
 H^{el}_{kn}(\bR)  c_n -
i\hbar \frac{{\bP}}{M} \cdot \bd^0_{kn}(\bR)  c_n \right)
\label{eq:etdse0}
\end{eqnarray}
That, at the end of this analysis, our conclusion was that the derivative coupling -- which is responsible for hopping between surfaces -- must be dressed if it is to be useful for surface hopping.
Given precedence in the literature\cite{bates:1958:etf,delos:1978:pra:etf,schneiderman:1969:pr:etf,winter:1982:pra:etf,errea:1994:etf,ohrn:1994:rmp:etf},  we referred to the extraneous terms in Eq. \ref{eq:etf}  as electron translation factors, and we argued that  in order to restore ``momentum  conservation,'' all that was required was that the derivative couplings be corrected by such electron \joeb{translation} factors (i.e. have them removed). Thus, we prescribed a correction to surface hopping: hop in direction $\bd^0$ rather than $\bd$ (following the decomposition in Eqs. \ref{eq:etf0}-\ref{eq:etf}):
 \begin{eqnarray}
\bP_{new} = \bP_{old} +\alpha {\bf d^0}     
\label{eq:rescale0}
 \end{eqnarray}

 At the time of  Ref. \citenum{fatehi:2011:dercouple},  we believed the analysis above solved surface hopping's momentum problem. Admittedly,  though, we always had several questions in the back of our mind.   
 \begin{itemize}
     \item 
First,  by removing the electron translational part of the derivative couplings,  we were able to restore linear momentum conservation; but what about angular momentum conservation?  That lack of conservation was still a problem. (And in the context of modern science,  had we focused on the angular momentum problem, we also would have recognized that simply removing the ETF component of $\bd$ would not be enough for us to make contact with systems with spins and degeneracy.) 
 
 \item Second, while Eq. \ref{eq:rescale0} restores linear momentum conservation during a hop, there was always an ambiguity about whether to implement  Eq. \ref{eq:etdse} or Eq. \ref{eq:etdse0}  for propagating the electronic amplitudes $c_k$. On the one hand,   when running surface hopping dynamics, seasoned practitioners never evaluate derivative coupling when calculating hopping rates\cite{shs:1994:protons,levine:2014:trivial,jain:2016:fast,lischka:2012:jcp_localized_diabatization}; instead, they note that   the Tully rate of hopping term is proportional to the time-derivative coupling:
 \begin{eqnarray}
     \bd_{nk} \cdot \bP/M = \left<\Phi_n \middle| \bnabla \Phi_k \right> \cdot \dot{\bR} = \left<\Phi_n \middle| \partial \Phi_k /\partial t \right>
     \label{eq:ddt}
\end{eqnarray}
 However, for adiabatic states $\Phi_n$ from Eq. \ref{eq:diagHBO}, the simplification in Eq. \ref{eq:ddt} holds if and only if we use $\bd$ (and not $\bd^0$); thus, Eq. \ref{eq:etdse} might appear the better choice. That being said, Eq. \ref{eq:rescale0} follows from  Eq. \ref{eq:etdse0}, and  only  Eq. \ref{eq:etdse0} 
 prevents an artificial population exchange arising from total translation motion (if we follow the consistency of Eq. \ref{eq:consistency}).
So neither equation of motion (Eq. \ref{eq:etdse} or Eq. \ref{eq:etdse0}) is perfect:  What to do?
 
 \item Third, one must wonder: can surface hopping be applied to the problem of hydrogen atom ionization? If so, how could that be possible if we were to remove  the only nonzero part of the derivative coupling?
 
\item  Fourth, given that the resolution to the momentum problem in FSSH appeared to arise from the phase of the electronic wavefunction, we always wondered what was the connection between the ETF component of the derivative coupling operator and the famous Mead-Truhlar-Berry phase\cite{meadtruhlar:1979:berry,berry:1984:berryphase}; several leading theorists at the time told us not to worry about this connection. They were wrong.   
 \end{itemize}
In retrospect, we should have also asked a fifth question, namely: what was the connection between  electron translation factors and Shenvi's PSSH formalism?  Was not $\hat{\tilde{H}}$ in  Eq. \ref{eq:newH} a Shenvi-like hamiltonian? At that time, however, we had become discouraged with PSSH for some of the reasons listed in Sec. \ref{sec:shenvi}. 
 The bottom line is that, in our excitement, we ignored several basic problems above that point to the need for a phase space electronic structure formalism;  sometimes people in a rush to claim victory miss the elephant in the room.





\subsection{The "Undefined" Problem of Spin  and Complex hamiltonians for Surface Hopping}
As far as we aware, the first chemists to focus on semiclassical surface hopping calculations of intersystem processes were Gonzalez and Persico and co-workers.\cite{gonzalez:2011:sharc,granucci:2012:fssh_spinorbit} In 2016,  one member of our research group began looking at problems of intersystem crossing.  Our motivation was that, because nonadiabatic dynamics are perennially expensive on account of the cost of electronic structure calculations and because we had our own implementation of code for CIS derivative couplings in Q-Chem\cite{qchem4},  we wondered whether we could speed up simulations of ISC by  including the spin-orbit coupling within the hamiltonian before diagonalization and then applying  the Hellmann-Feynman theorem to extract cheap derivative couplings efficiently.  We were also curious about whether we could see magnetic field effects during ISC simulations\cite{cederbaum:1997:ijqc:magnetic_fied_thoughts}, so we decided to include magnetic fields in our calculations. In retrospect, we were clearly biting off more than we could chew at the time.

The problem that emerged, of course, is that in the presence of a magnetic field, the electronic hamiltonian becomes complex.
(Note that, even though the SOC is complex-valued, Mead showed that in the absence of a magnetic field, the hamiltonian can still be made entirely real for a system with an even number of electrons by using the correct basis.\cite{mead:1979:noncrossing}
 The complex part of the problem arises only in the presence of a magnetic field for a system with an even number of electrons.)   Thus, during the course of our calculations above, we began calculating derivative couplings that were complex -- which immediately raised our eyebrows. 
After all, how does one run surface hopping trajectories in the presence of complex-valued derivative couplings?  Eq. \ref{eq:rescale} is simply meaningless: a classical momentum  cannot be complex-valued.  Thus, the basic question above permanently changed the course of research in our laboratories.

In order to aquire some basic intuition for the problem,  several of the present authors immediately began to run exact and surface hopping calculations for a model hamiltonian of the following form\cite{miao:2019:fssh:complex},
\begin{eqnarray}
    \hH_{flat,2D} = \frac{\hP_X^2}{2M} + \frac{\hP_Y^2}{2M}+ \left(
\begin{array}{cc}
\cos(\theta(\hX)) & \sin(\theta(\hX)) e^{iW\hY} \\
\sin(\theta(\hX))e^{-iW\hY} & -\cos(\theta(\hX))
\end{array}
    \right)
    \label{eq:2d:flat}
\end{eqnarray}
 which is a clear generalization of  Eq. \ref{eq:Hflat} above from one dimension $(X)$ to two nuclear dimensions $(X,Y)$.

Over the course of several years, we tried multiple approaches for rescaling the momentum and compared against exact simulations.  After several partially satisfactory attempts\cite{miao:2019:fssh:complex,wu:jcp:2021:first_attempt_complex,bian:jctc:2022:first_attempt}, we finally realized\cite{wu:2022:pssh,bian:2022:pssh}  there was only one meaningful means to solve the dynamics of Eq. \ref{eq:2d:flat} semi-classically.  Namely, one needed to first apply a unitary transform and generate an electronic hamiltonian that depended on $\bR=(X,Y)$ and $\bP=(P_X,P_Y)$:

\begin{eqnarray}
U &\equiv& \left( 
\begin{array}{cc}
e^{-iWY} & 0 \\ 0 & 1
\end{array}
\right)
\\
    H' &=& U \hH^{flat}_{2D}  U^{\dagger}  
    \nonumber
    \\
    \label{eq:ps_flat}
    &=&  \frac{\hP_X^2}{2M} + \frac{1}{2M}
    \left(
\begin{array}{cc}
(\hP_Y + \hbar W)^2 &   \\
0& \hP_Y^2
\end{array}
    \right) +
    \left(
\begin{array}{cc}
\cos(\theta(\hX)) & \sin(\theta(\hX))  \\
\sin(\theta(\hX))& -\cos(\theta(\hX))
\end{array}
    \right)
    \label{eq:UHU}
\end{eqnarray}

The result of the unitary transform in Eq. \ref{eq:UHU} is that, unlike Eq. \ref{eq:2d:flat},  the  hamiltonian has become completely real -- but at the expense of introducing new dependence on the nuclear momentum, i.e. a vector potential. In other words, just as in Eq. \ref{eq:Hshenvi} above, the electronic hamiltonian in Eq. 
\ref{eq:UHU} is parameterized by both $\bR$ and $\bP$, and so it would 
appear natural to follow Shenvi's phase space surface hopping PSSH approach and run dynamics along phase space potential energy surfaces.
Indeed, when we performed such calculations, we found\cite{wu:2022:pssh,bian:2022:pssh} that the results were quantitatively accurate as compared with exact simulations -- far, far more accurate than was possible with FSSH --  a fact which made us return to Shenvi's phase space surface hopping approach for inspiration when considering how to include spin and magnetic fields.

\section{Berry Curvature Distractions }
\label{sec:berry}

 At this point, another digression is in order.  When one runs dynamics along a single Born-Oppenheimer adiabat for the hamiltonian in Eq. \ref{eq:2d:flat}, one finds that every trajectory bends according to the sign of $W$ and the nature of the adiabat (upper or lower).   Indeed, Eq. \ref{eq:ps_flat}  makes clear there is an intrinsic magnetic field (with vector potential $W$) along one diabatic component state of this problem.
 The phenomenon above  can be proven  quite generally.  If for any hamiltonian, one wants to run classical dynamics along surface $E_0$,  Berry and Robbins\cite{berry:1993:royal:half_classical} proved that dynamics with a Born-Oppenheimer force $\bF_{BO}=-\bnabla E_0$  should be augmented by a magnetic field of the form
 \begin{eqnarray}
 \bF_B = 2\hbar \sum_{j \ne 0} \mbox{Im}\left(  \bd_{0j} \left( \bd_{j0} \cdot \bP/M \right) \right)  
 \label{eq:berryforce}
 \end{eqnarray}
This so-called ``Berry Force'' is exactly what forces single-state trajectories to curve for the flat adiabats from Eq. \ref{eq:2d:flat}. While one can derive Eq. \ref{eq:berryforce} in many different ways, heuristically the answer is equivalent to assuming that $d_{00}$ in Eq. \ref{eq:Hshenvi} be treated like an external vector potential; Eq. \ref{eq:berryforce} is then the corresponding magnetic field. Note that $\bF_B = \bP/M  \times\left(\nabla \times i \hbar \bd_{00}\right) $.

One immediate and conceptually appealing aspect of running calculations with a Berry force is that the total momentum of a molecular system is changed (relative to dynamics without such a force).  In fact, from a BO perspective, one can argue that $\sum_I \bd^I_{00}$ represents the electronic momentum along surface 0. The argument for such an interpretation is as follows\cite{littlejohn:2023:jcp:angmom,littlejohn:2024:jcp:moyal}:  for quantum BO dynamics along a single surface, consider the total momentum operator operating on a combined nuclear + electronic wavefunction, $\chi(\bR) \Phi(\br;\bR):$
\begin{align}
& \left( \frac{\partial }{\partial \br} +
    \sum_I \frac{\partial }{\partial \bR_I}\right)
    \chi(\bR) \Phi(\br;\bR) = \nonumber \\
   &= \; \; \;  \; \; \; \; \left(    
    \sum_I \frac{\partial }{\partial \bR_I} \chi(\bR)
    \right) \Phi(\br;\bR)  + 
    \left(      \left( \frac{\partial }{\partial \br} +
    \sum_I \frac{\partial }{\partial \bR_I }\right)
    \Phi(\br;\bR) \right)    \chi(\bR)
\end{align}
Now, if one makes the natural choice of gauge that the electronic wavefunction is defined relative to  the nuclear positions,  this  choice implies that $\left( \frac{\partial }{\partial \br} +
    \sum_I \frac{\partial }{\partial \bR_I }\right)
    \Phi(\br;\bR) =0$.
Thus, within this interpretation, $\left(\sum_I \bP_I\right) \chi(\bR) \equiv  -i \hbar \left( \sum_I \partial/\partial \bR_I \right) \chi(\bR)$ represents the operation of the {\em total} nuclear+electronic momentum. 
Therefore, given that $\sum_I  (\bP_I - i\hbar \bd^I_{00})^2/(2M_I)$ represents
the nuclear kinetic energy for state 0 in Eq. \ref{eq:Hshenvi}, it follows that  $i \hbar \sum_I \bd^I_{00}$ must represent the electronic momentum (see Eq. \ref{eq:early_etf}).

In direct line with this interpretation, one can show that classical dynamics along a BO surface {\em with a Berry force} do conserve the {\em total, nuclear+electronic} momentum, both linear and angular. This result can be shown\cite{xuezhi:2023:total_ang_bomd} from the basic fact that the BO potential energy surface is both translationally invariant and isotropic
(and extensions also hold for dynamics in external magnetic fields as well\cite{helgaker:2021:jcp:dynamics_in_magnetic_field,helgaker:2022:jcp_conservation_laws_magnetic_field}).
Interestingly, these conclusions do not hold if one throws out the ETF part of the derivative coupling (from Eq. \ref{eq:etf} above).  In other words, one cannot disentangle the Berry force (or Berry phase) from the ETF component of the derivative coupling; the two objects are very related and are necessary for understanding momentum conservation (and this momentum conservation must hold everywhere in configuration space, not just near conical intersections).

 Because of the very fundamental properties listed above, our research group spent several years and published several papers aiming to build a surface hopping algorithm that combined Born-Oppenheimer dynamics  with Berry forces along individual adiabatic surfaces\cite{miao:2019:fssh:complex,wu:jcp:2021:first_attempt_complex,bian:jctc:2022:first_attempt}.  We were originally very hopeful that such a merger would be a powerful new tool for theoretical chemistry\cite{bian:2021:perspective}.  That being said, at the end of the day, we came to the conclusion that such a merger of surface hopping and Berry forces was not healthy or promising  for several reasons.
First,  empirically for the model problem in Eq.  \ref{eq:2d:flat}, we found that phase space surface hopping strongly outperformed standard surface hopping with Berry forces (and the latter was often qualitatively incorrect). For the skeptic, simply compare the results of Refs. \citenum{wu:2022:pssh,bian:2022:pssh} vs those of Refs. \citenum{wu:jcp:2021:first_attempt_complex,bian:jctc:2022:first_attempt}!  Second, a major limitation of Berry forces is that they are only applicable for non-degenerate surfaces; as such, any algorithm predicated on Berry forces will never be applicable for spin systems  with odd numbers of electrons due to Kramers degeneracy.  Third,  consider dynamics for a system with an even number of electrons. As discussed above\cite{mead:1979:noncrossing},  the total hamiltonian can always be made real -- with real value derivative couplings.  Thus, for a system with an even number of electrons, the Berry force will necessarily vanish and we must predict zero electronic momentum (even if there is physically electron transfer). To make matters more complicated, calculations  suggest that the Berry force as computed with an approximate, symmetry-broken solution are in fact meaningful\cite{bian:2022:jcp:meaning} (even though the exact Berry force would be zero).  The bottom line is that, for problems with spin, there is no unique Berry force for a single adiabatic surface but rather a matrix of Berry forces (i.e. the nonabelian Berry curvature) and the latter is compatible with Ehrenfest dynamics\cite{takatsuka:2005:jcp,krishna:2007:ehr_plus_berry,coraline:2024:jcp:ehrenfest_conserve} -- but is not easily compatible with a surface-hopping approach. 

In the end, although we spent several years working on Berry curvature -- a field which is also called quantum geometry among physicists\cite{polkovnikov:2017:quantum_geometry} --  we came to believe that working with Berry forces on top of Born-Oppenheimer potentials is not a productive means to many nonadiabatic phenomena, especially non-adiabatic dynamics with spin.  We concluded that a more promising avenue is to invoke a phase space electronic structure approach, with electronic hamiltonians parameterized by $\bR$ and $\bP$ in the spirit of Eq. \ref{eq:Hshenvi} (but necessarily updated). If necessary, quantum geometry and Berry curvature effects can be added on top of this new reference.

\section{The Emergence of  a Robust Phase Space Approach}

At this stage,  many of this article's  authors  made the conscious decision to develop and pursue a phase space view of electronic structure in the spirit of  (but not equivalent to) Eq. \ref{eq:Hshenvi}.  Naturally, the most difficult question is: how to construct the relevant phase space electronic hamiltonian and specifically the correct momentum coupling term.  As discussed in Sec. \ref{sec:shenvi} above, we wanted to avoid  working directly with the $\bd \cdot \bP/M$ term, as that approach leads to many numerical and conceptual problems.  For the model problem in Eq. \ref{eq:2d:flat},  the alternative was clear: one could perform a unitary transform and isolate an obvious $\hH_{PS}$, where  the novel momentum-coupling was obvious (i.e. $PW/M$ in Eq. \ref{eq:ps_flat}).    But how should we build the momentum-coupling in general for {\em ab initio} molecular or materials problem where there was no $W$ operator? More concretely, if we wanted to build a phase space hamiltonian of the form
 \begin{eqnarray}
 \label{eq:HPSG0}
     \tilde{H}_{PS} \stackrel{?}{=}  \frac{(\bP - i\hbar \tbGamma)^2}{2M} + \hH^{el}_{diag} 
 \end{eqnarray}
 where $\tbGamma$ was a stable one-electron approximation of the derivative coupling, what should $\hbGamma$ be? 
 
 Before answering this question, it is worth noting that because $\hH^{el}_{diag} = \hU^{\dagger} \hH \hU$ for the matrix of eigenvectors $\hU$,  $\tilde{H}_{PS}$ in Eq. \ref{eq:HPSG0} can be transformed unitarily into 
 \begin{align}
     \hH_{PS}(\bR,\bP) & = \hU \tilde{H}_{PS} \hU^{\dagger} \\
     & = \frac{(\bP - i\hbar \hbGamma(\bR))^2}{2M} + \hH_{el} 
     \label{eq:HPSG1}
 \end{align} 
where $\hbGamma \equiv \hU \tbGamma \hU^{\dagger}$.  In other words, the modern phase space approach to electronic structure is very different from Eq. \ref{eq:Hshenvi} insofar as we perform only one  diagonalization (i.e. we diagonalize  $\hH_{PS}$ in Eq. \ref{eq:HPSG1}) rather than perform two diagonalizations (i.e. Shenvi's approach would have us  diagonalize $\hH_{el}$ in Eq. \ref{eq:diagHBO} first and then $\hH_{Shenvi}$ in Eq. \ref{eq:Hshenvi} second). The former approach is much more computationally efficient than the latter.

Let us then return to the question of how to construct $\hbGamma$ (which is much more convenient than $\tbGamma$).
After a great deal of time to consider the problem,  eventually it dawned on us that one approach for building $\hbGamma$ had been staring at us right in the face all along: namely, why not construct $\hbGamma$ from the ETF part of Eqs. \ref{eq:etf0}-\ref{eq:etf}?  In other words, rather than throw out the ETFs from the derivative coupling when performing a surface hopping calculation (see Sec. \ref{sec:hop_etf}),  why not  include that same ETF component of the derivative coupling as an electron-phonon coupling and set 
\begin{eqnarray}
    \hbGamma \stackrel{?}{=} \hat{\bd}^{ETF}
    \label{deqetf}
\end{eqnarray}
for a phase space electronic structure calculation? After all, $\hat{\bd}^{ETF}$ is an effective, one-electron approximation for the electronic momentum and is easily computable as well. 

But how could we prove that such an ansatz  was   physical and would be helpful?    Besides comparing with experimental observables, we had several theoretical and/or computational tools at our disposal. First, one of the goals of a phase space electronic structure approach is to ensure that dynamics  conserve the total linear and angular momentum of an isolated system.
To that end,  in  Ref. \citenum{yanzewu:2024:jcp:pssh_conserve}, we   showed that dynamics along an eigensurface of $\hH_{Shenvi}$ in Eq.  \ref{eq:Hshenvi} did indeed conserve linear and angular momentum. Furthermore, Ref. \citenum{yanzewu:2024:jcp:pssh_conserve} demonstrated  that conservation arises from  four different symmetries that had been anticipated by Littlejohn {\em et al.} \cite{littlejohn:2023:jcp:angmom,littlejohn:2024:jcp:moyal}:
\begin{eqnarray}
    -i\hbar\sum_{I} {\bm d}^I_{jk} + \bra{\Phi_j}\bm{\hat{p}} \ket{\Phi_k} &=& 0,\label{eq:dconstrain1a}  \\
    \sum_I \bm \nabla_{I} {\bm d}^J_{jk}  &=& 0, 
    \label{eq:dconstrain2a}\\
    -i\hbar\sum_{I}{\bm X}_{I} \times {\bm d}^I_{jk} + \bra{\Phi_j} \bm{\hat{l}} + \bm{\hat{s}} \ket{\Phi_k} &=& 0,\label{eq:dconstrain3a}\\
     -\sum_I \left({\bm X}_I \times \bm \nabla_{I} {d}^{J\beta}_{jk}\right)_{\alpha} &
     =&  \sum_{\gamma} \epsilon_{\alpha \beta \gamma} d^{J \gamma}_{jk},\label{eq:dconstrain4a} 
\end{eqnarray}

Thus, our next goal was to check whether or not $\bd^{ETF}$ in Eq. \ref{eq:etf} alone  satisfies these same equations. A simple calculation revealed that $\bd^{ETF}$ satisfies Eqs. \ref{eq:dconstrain1a} and \ref{eq:dconstrain2a}, which are  
needed to conserve linear momentum; however, $\bd^{ETF}$ does not satisfy Eqs. \ref{eq:dconstrain3a} and \ref{eq:dconstrain4a}, which are needed to conserve angular momentum.  Thus, in Refs. \citenum{athavale:2023:erf}-\citenum{tian:2024:jcp:erf}, we immediately realized that, in order to conserve both linear and  angular momentum,  one must also develop and include {\em electron rotation factors} to complement the ETFs. The final result is that we can write
\begin{eqnarray}
\label{gammappp}
    \hbGamma = 
    \hbGamma' +
    \hbGamma''  +
    \hbGamma''' 
\end{eqnarray}
where $\hbGamma'$ is required to conserve linear momentum, $\hbGamma''$ is required to conserve orbital angular momentum, and $\hbGamma'' + \hbGamma'''$
is required to conserve the total orbital plus spin angular momentum.

Second, noting the argument above in Sec. \ref{sec:berry} that Berry forces also conserve the total momentum (but that the electron momentum is actually zero very often), it is crucial to prove that PSEST predicts accurate, {\em nonzero} electronic momentum. To that end, the simplest calculation possible is to perform two electronic structure calculations on a molecule at two slightly different nuclear positions and then to calculate the change in the electronic position.  If one divides the latter by the change in nuclear position, one should recover the ratio between the electronic momentum and the nuclear momentum.
In other words, one must check that:
\begin{eqnarray}
\label{eq:nafie}
    m_e \frac{d\left< \Phi_{BO} \middle | \hbr_e \middle | \Phi_{BO} \right>}{dt} \approx \left< \Phi_{PS} \middle | \hbp_e \middle | \Phi_{PS} \right> \Leftrightarrow 
    \frac{m_e}{M_I} \frac{\partial \left< \Phi_{BO} \middle | \hbr_e \middle | \Phi_{BO} \right>}{\partial R_I} \approx \frac{\partial \left< \Phi_{PS} \middle | \hbp_e \middle | \Phi_{PS} \right>}{\partial P_I}
\end{eqnarray}
Indeed, Nafie effectively showed long ago\cite{nafie:1983:jcp:el_momentum,coraline:2024:jcp:pssh_conserve} that Eq. \ref{eq:nafie} was satisfied exactly if one used Shenvi's hamiltonian in Eq. \ref{eq:Hshenvi}, and so one would hope that the same result would hold approximately using $\hbGamma$.  The results of this calculation were presented in Ref. \citenum{coraline:2024:jcp:pssh_conserve}
and are a strong endorsement that our
approximate $\hbGamma$ was satisfactory.

Third, one of the major shortcoming of Tully's FSSH is that a global translation of a molecular system was able to induce an electronic transition between bound states. As mentioned above in Sec. \ref{sec:hop_etf}, this premise is nonphysical: a hydrogen atom in a $2p_x$ does not hop down to the $1s$ state spontaneously. By contrast, if one now runs surface hopping on phase-space surfaces (rather than BO surfaces), one can indeed show\cite{bian:2024:pssh_translations_rotations} that these fictitious transitions in FSSH are entirely eliminated within phase space surface hopping (PSSH); of course, these transitions are also eliminated by removing the ETFs of the derivative couplings within FSSH. Perhaps more poignantly, however, one can also show that the hopping rate is much reduced under PSSH relative to FSSH upon a global rotation; more precisely, for a global rotation, the hopping rate within FSSH is proportional to $\hbar$ while the hopping rate within PSSH is proporational to $\hbar^2$ (and effectively negligible). Thus, PSSH has strongly reduced the hopping rate, which is a clear advantage of the new method.

Fourth and lastly, Ref. \citenum{duston:2024:jctc_vcd}  demonstrated that, when applied with a large enough atomic orbital basis, the {\em ad hoc} ansatz that $\hbGamma = \bd^{ETF}$ does in fact  recover a vibrational circular dichroism spectrum for several small chiral molecules (e.g. \joeb{partially deuterated} oxirane).  This particular success  provided immense motivation for the present authors to continue searching for improved phase space electronic structure hamiltonians.

\subsection{A Basis Free $\hbGamma$: A Major Step Forward}

The next major step forward came in 2024 when we began to run PSEST calculations over larger and larger   atomic orbital sets, and we consistently found that the results would depend strongly on basis sets.\cite{zhentao:2024:jcp:vcd_basis_free,coraline:basisfree:2025}   Were these results a signal that diagonalizing $\hH_{PS}$ was intrinsically unstable, or rather a problem of setting $\hbGamma = \hbd^{ETF}$ ?  Eventually, we traced the problem down to our choice of $\hbGamma$: for a set of atomic orbitals $\left\{ \chi_{\mu}\right\}$ if one visualizes $f(\br) \equiv \sum_{\mu \nu} \bd^{ETF}_{\mu \nu} \chi_{\mu}(\br) \chi_{\nu}(\br)$ from Eq. \ref{eq:etf}, one finds that the matrix does not converge with basis.  Indeed, the experience of writing Ref. \citenum{duston:2024:jctc_vcd} had been  unnecessarily challenging because the construction of any second-quantized operator in an atomic non-orthogonal basis is always ugly. Thus, we realized that what was needed was an expression for $\hbGamma$ in terms of $\hbr,\hbp,\hbR,\hbP$, i.e. an expression that did not rely on any atomic orbital basis.

Our solution to this conundrum\cite{coraline:basisfree:2025} was to replace equations \ref{eq:dconstrain1a}-\ref{eq:dconstrain4a} (i.e. the constraints for the derivative couplings) with analogous constraints for $\hbGamma$ in operator form:
\begin{eqnarray}
    -i\hbar\sum_{I}\hbm{\Gamma}_{I} + \hat{\bm p} &=& 0,\label{eq:constrain1}  \\
    \Big[-i\hbar\sum_{J}\pp{}{\bm{R}_J} + \hat{\bm p}, \hbm{\Gamma}_I\Big] &=& 0,\label{eq:constrain2}\\
    -i\hbar\sum_{I}{\bm R}_{I} \times \hat{\bm \Gamma}_{I} + \hat{\bm l} + \hat{\bm s} &=& 0,\label{eq:constrain3}\\
     \Big[-i\hbar\sum_{J}\left(\bm{R}_J \times\pp{}{\bm{R}_J}\right)_{\gamma} + \hat{l}_{\gamma} + \hat{s}_{\gamma}, \hat{\Gamma}_{I \delta}\Big] &
     =& i\hbar \sum_{\alpha} \epsilon_{\alpha \gamma \delta} \hat{\Gamma}_{I \alpha}. \nonumber \\
     \label{eq:constrain4} 
\end{eqnarray}
One can then construct  $\hbGamma$ in a basis free fashion by partitioning space through a partition of unity $\left\{ \theta_1, \theta_2, \ldots, \theta_{N_A} \right\}$ (where $N_A$ is the number of atoms).  Here, a partition of unity is a set of functions that divide up space smoothly between different atoms.  The easiest form for $\theta_I$ is of Hirschfelder form:
\begin{align}
    \theta_I(\br) = \frac{\exp(-(\br - \bR_I)^2/\sigma^2)} {\sum_J \exp(-(\br - \bR_J)^2/\sigma^2)}
    \label{eq:theta}
\end{align}
Thus, $\sum_I \theta_I(\hbr) = 1$ but $\theta_I(\br)$ is nonzero only when $\br$ is near atom $\bR_I$.

Armed with this (non-unique) partitioning, one can replace the ETF ansatz in Eq. \ref{deqetf} with the function $\left(\hbp \theta_I + \theta_I \hbp\right)/2$:

\begin{centering}
\begin{eqnarray}
\begin{array}{ccc}
\Gamma'^I_{\mu \nu} = \underbrace{ \sum_{p q}  \frac{1}{2i\hbar} 
\left( \left< \mu \middle | \frac{\partial \nu}{\partial R_I} \right>
-
\left( \left< \frac{\partial \mu}{\partial R_I} \middle | \nu   \right> \right)
\right)
C_{\mu p} C_{\nu q}  a_p^{\dagger} a_q }   & \Rightarrow &  \Gamma'^I_{\mu \nu} = \underbrace{\frac{1}{2i\hbar} \left(\hbp \theta_I + \theta_I \hbp\right)_{\mu \nu}}   \\
    \mbox{atomic orbital basis} & & \mbox{basis-free}
\end{array}
\label{eq:newgamma'}
\end{eqnarray}
\end{centering}
A similar construction can be made for the ERFs as well.  The full, basis-free expression that emerged for $\hbGamma$ can be found in Eqs. 37-47 in Ref. \citenum{xuezhi:cpr:review:2026}, and are not repeated here. Benchmarks of charge density and VCD spectra confirmed\cite{zhentao:2024:jcp:vcd_basis_free} that this basis-free choice of a phase space electronic hamiltonian provided vastly smoother electronic densities with stable complete basis set limits.

\subsection{Wigner transforming and Littlejohn-Flynn Theory}

At this point, one might think that our work is done: we have presented a historical narrative framing the construction of a new phase space electronic Schrodinger equation that can be solved and was tested against VCD experiments\cite{zhentao:2024:jcp:vcd_basis_free}. That being said, however, beyond VCD spectra and a consistency check for electronic momentum, we have not mentioned any  other theoretical benchmarks to compare against. How confident can we be in our choice of the $\hbGamma$ operator in Eq. \ref{eq:newgamma'}? And if so, how should we pick the parameters in $\theta_I$ in Eq. \ref{eq:theta}? More generally, phase space electronic structure theory cannot be complete without a mapping of the electronic operator $\hH_{PS}$ to the full nuclear-electronic Schrodinger equation: without such a mapping, we would be unable to be correctly move nuclei along  surfaces. And in practice, without such a mapping, we would not be able to compare our results rigorously against exact quantum solutions for nonadiabatic problems. This state of affairs was very frustrating for us because generations of quantum dynamicists have been able to  test BO theory on model problems\cite{tully:faraday:fssh,coker:2008:iterative,miller:2007:ananth} and find its failures and gain intuition, and we similarly wanted to test PS theory to see how it fared during analogous model-problem benchmarking.

Of course the major obstacle preventing a more through benchmarking of phase space electronic structure theory was the fact  that the nuclei are treated as classical variables -- not quantum mechanical operators. After all, if one thought of $\bP$ as $\hbP$ in Eq. \ref{eq:Hshenvi}, then Eq. \ref{eq:diagHShenvi} is the full electron-nuclear Schrodinger equation; Eq. \ref{eq:HPSG1} would also be impossible to solve (and would not even be the correct Schrodinger equation either!).  To that end, in Ref. \citenum{bian:2025:jctc:wigner_vibrations} we made our first attempt to build a rigorous,  fully quantum  phase space electronic structure theory for comparison  against exact quantum results.  Our target was  a one-dimensional model problem of hydrogen motion introduced by Borgis {\em et al}\cite{borgis:model:xuezhi:gross:wigner:cpl:2006} and studied subsequently by Gross {\em et al}\cite{gross:2017:prx:born_oppenheimer_mass}.   Our approach was to first solve the ground state PES in phase space, and then second requantize the nuclei using a Weyl transform as in Eq. \ref{eq:weyl}. Our results were quite revealing: not only were we able to recover improved vibronic energies, we were also able to capture very accurate other matrix elements – most notably, electronic momentum (which BO completely neglects).
For those interested in exploring PS electronic structure methods for themselves, reproducing Fig. 2 of Ref. \citenum{bian:2025:jctc:wigner_vibrations} is a good start.

At this juncture, a practical electronic structure theorist might again want to declare victory: we had shown that a PS approach could meaningfully outperform BO  theory -- not just for electronic observables but also for nuclear observables  on a single potential surface.  But for a quantum dynamicist interested in surface hopping, the next question is obvious: how should we piece together these dynamics on different electronic surfaces so as to compare with model problems? After all, as far as nuclear dynamics, a simple Berry force approach can sometimes be sufficient  for a single surface, but as discussed in Sec. \ref{sec:berry},  bigger problems arise when combining dynamics on different surfaces.  And whereas the BO framework is well-defined for multiple surfaces (e.g. so that one can converge systematically to an exact solution), there was no such rigorous approach within phase space electronic structure theory.


 To that end, in Ref. \citenum{xinchun:2025:wigner_one_state}, we realized we could find a rigorous framework for phase space electronic structure theory (so as to tackle nonadiabatic eigenstate problems) by applying Littlejohn-Flynn theory\cite{littlejohn:flynn:1991:pra:coriolis}.
 To understand how LF theory operates in this case,
imagine that we have diagonalized $\hH_{PS}$ at every $\bR$ and $\bP$ point and  generate a matrix of electronic eigenvectors and eigenvalues:

\begin{eqnarray}
\label{eq:diagHPS}
    \hH_{PS} = L(R,P) \Lambda(R,P) L^{\dagger}(R,P)
\end{eqnarray}

Eq. \ref{eq:diagHPS} is  not a true diagonalization of the total hamiltonian; it is a diagonalization of only the electronic degrees of freedom. Following up on Eq. \ref{eq:Wstar} above, if we want exact nuclear-electronic eigenvalues, what we really seek is a matrix decomposition of the form:

\begin{eqnarray}
\label{eq:diagHPS_star}
    \hH_{W} = J(R,P) * \Lambda'(R,P) * J^{\dagger}(R,P)
\end{eqnarray}

The Littlejohn-Flynn  approach is thus to build $\hJ$ as a perturbation of  $\hL$  and to build $\hLambda'$ as a perturbation of $\hLambda$ --- while also insisting that $\hJ$ is unitary in the joint-nuclear electronic space (unlike $\hL$, which is unitary only in the electronic space).  This approach was successful in Ref. \citenum{xinchun:2025:wigner_one_state} and  demonstrated  that one could generate a  completely rigorous view of phase space electronic structure and converge  to the exact answer over many electronic states for a nonadiabatic problem.  Admittedly, Littlejohn-Flynn theory can be painful to implement in practice and in many cases, if one seeks the exact solution, the BO framework is still easier. That being said, for most hamiltonians, if one simply wants a great deal of accuracy (but not the exact solution), we could now rigorously justify a phase space electronic structure approach.

\subsection{Curl-Free Conditions and Exact Gauge potential}
\label{sec:curl}

As of early 2026,  the most recent theoretical change to our understanding of phase-space electronic structure theory comes from a very simple calculation of the nonabelian curl of $\hbGamma$ for a set of atoms restricted to be in one dimension. For this unique case\cite{zain:inprogress}, one can show that  the $\hbGamma$ operators satisfy the condition:
\begin{eqnarray}
    \frac{\partial \hbGamma^A}{\partial R_B} -  \frac{\partial \hbGamma^B}{\partial R_A} - \left[ \hbGamma^A, \hbGamma^B  \right]=0
    \label{eq:curl}
\end{eqnarray}
The left hand side of Eq. \ref{eq:curl} is known as the non-abelian curl. Because the right hand side of Eq. \ref{eq:curl} is zero, 
it follows that we can write:
\begin{eqnarray}
\label{eq:defUG}
    \hbGamma^A = \hU^{\dagger}\frac{\partial \hU}{\partial \bR_A}
\end{eqnarray}
for some electronic unitary matrix parameterized by $\bR$, i.e. $\hU(\bR)$.

From the result above, one can show rigorously that in one dimension, the phase space electronic hamiltonian is an exact representation of the total nuclear electronic hamiltonian. In particular, it follows that
\begin{eqnarray}
    \hH_{tot} = \hU^{\dagger}  \bigg( \; W^{-1}\left(\hU  \hH_{PS} \hU^{\dagger} \right)  \biggr) \hU
    \label{eq:HtotfrompsU}
\end{eqnarray}
for the $\hU$ defined in Eq. \ref{eq:defUG}.
Eq. \ref{eq:HtotfrompsU} should be compared against the standard quantum chemistry expression,
\begin{eqnarray}
    \hH_{tot} = W^{-1}\left( \hH_{BO} \right)  
    \label{eq:HtotfromBO}
\end{eqnarray}
where $\hH_{BO}$ is the Born-Oppenheimer hamiltonian.

Thus, in one dimension, one can view phase space electronic structure theory as a novel preconditioner of the original nuclear-electronic hamiltonian. Whereas the Born-Huang framework arises from taking $\hbGamma =0$ and $\hU = \hat{I}$, phase space electronic structure theory in one dimension asks if there is a better preconditioner $\hU$ that can be applied -- one which both  leads to momentum conservation along a single surface and produces more accurate spectra. Again, there is no loss of information in this ansatz provided that the $\hbGamma$ in Eq. \ref{eq:HPSG1} satisfies the curl condition in Eq. \ref{eq:curl}. 

Alas, however, the curl condition does not hold for any meaningful $\hbGamma$ (at least, any that we have tested) in more than one dimension. In particular, we recently showed\cite{mansi:nadine:moody_shapere_wilczek} that, for a diatomic molecule,  our proposed $\hbGamma$ performs remarkably well while also satisfying and generalizing the Moody-Shapere-Wilczek gauge condition\cite{moody:shapere:wilczek:1986} -- whose curl includes a magnetic monopole (and is far from zero). Thus, the question remains as to how much physics (within PSEST) can be captured by a trivial gauge and how much requires a non-trivial gauge, a question that will need a great deal of further investigation to fully work out. In practice, we can expect that phase space electronic structure theory will give us much larger corrections in three dimensions relative to one dimension.


\section{The Future and Open Questions}

\subsection{Successful Experiments and  Possibilities for Future Material Science  and Magnetic Field Chemistry Applications}

We have now come to the present day. Thus far, we have explained the theory and the origins of PSEST.  Besides the model problems explored above, we note that the theory above has already  been tested on a broad range of experimental observables as applicable to systems with both degenerate and non-degenerate electronic structure.\cite{duston:2024:jctc_vcd,zhentao:2024:jcp:vcd_basis_free,linqing:2026:lambda_doubling,linqing:2026:spin_rotation,coraline:2026:roa} 

Regarding molecules with non-degenerate electronic ground states, the first set of experiments involves the spectroscopy of closed shell molecules within a chiral nuclear environment.  Here, there is no inversion center and the molecule/materials will be sensitive to the direction of an applied external magnetic field.  Moreover, because an external magnetic field   operates only on moving (not static) charges and because  BO does not predict the correct electronic momentum in the presence of nuclear motion, one must expect that BO theory will fail to quantitatively describe nuclear vibrations in a magnetic field.  By contrast, if one can construct a proper PSEST in the presence of a magnetic field\cite{bhati:2025:jpca:magnetic_part1,bhati_magnetic_part2}, one would expect that PSEST would perform well in such a case.  To that end, in Refs. \citenum{zhentao:2024:jcp:vcd_basis_free,coraline:2026:roa}, we indeed showed that PSEST is able to recover vibrational circular dichroism and Raman optical activity in both of these situations.  These tests represent critical tests for wavefunctions parameterized by both $\bR$ and $\bP$, and so far PSEST has risen  to the challenge for our initial choice of $\hbGamma$. 

\begin{figure}
    \centering
    \includegraphics[width=0.8\linewidth]{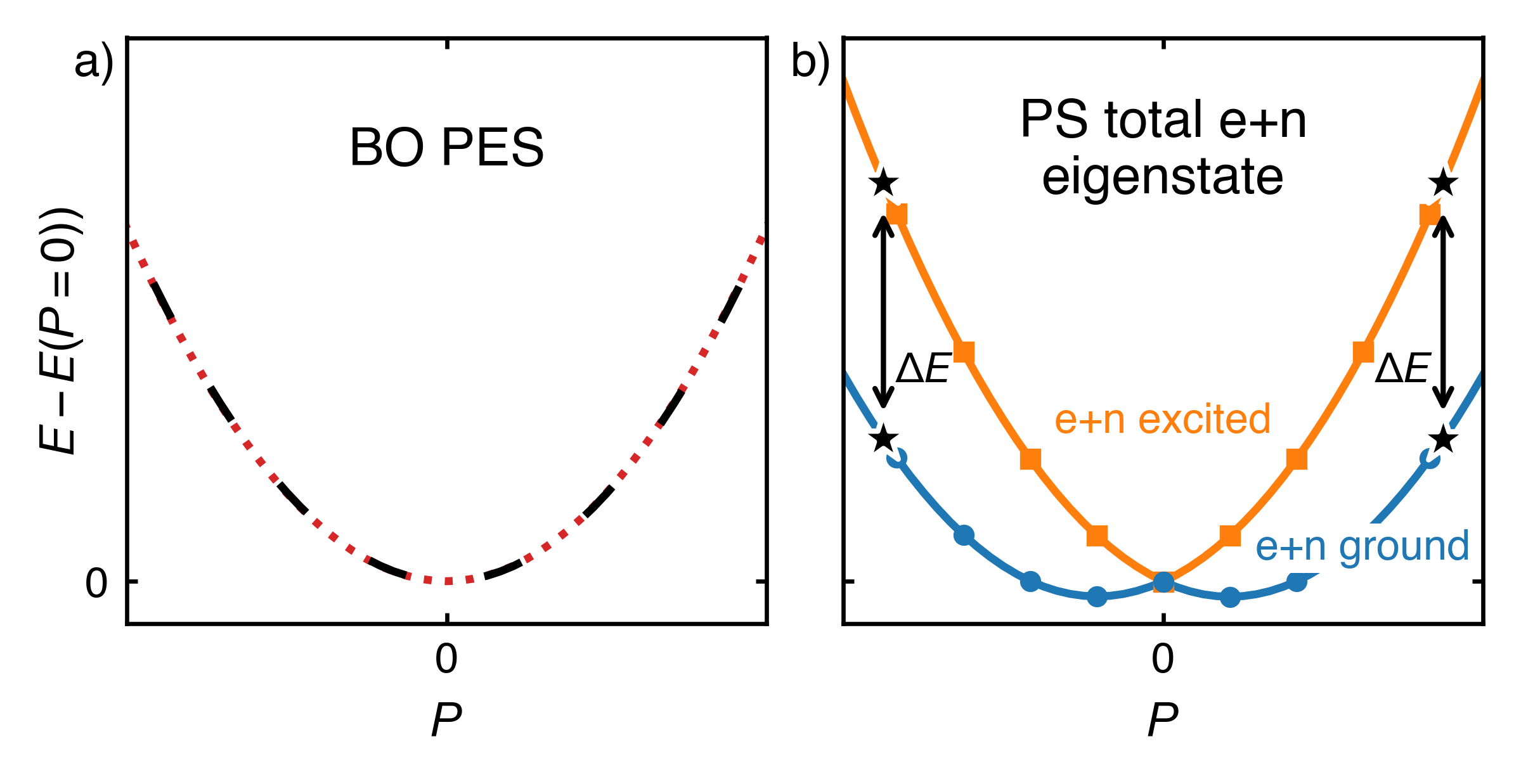}
    \caption{(a) Within BO theory, one separates the kinetic and potential energies for a molecule. Thus, for a system in an electronic doublet of (i.e. a set of Kramers pairs), the total energy is always degenerate as a function of nuclear momentum. (b) That being said, within PS theory, by accounting for noninertial frame effects and momentum coupling,  the  degeneracy at $\bP=0$  is removed  for $\bP \ne 0$ when time-reversibility is lost.  The energy gap between the resulting two states corresponds to $\Lambda$ doubling in diatomics (like NO) or spin-rotation coupling in small molecules and is experimentally measurable.  }
    \label{fig:kramers}  
\end{figure}

Regarding systems with electronic degeneracy or near degeneracy, the second set of experiments focuses on open-shell molecules  with spin degrees of freedom.  \joeb{To understand the interesting effects one can study with PSEST, note that (i) PSEST does preserve time-reversal symmetry for the joint nuclear-electronic problem. Namely, if $\hH_{PS}(\bR,\bP) \ket{\Psi(\bR,\bP)} = E_{PS}(\bR,\bP) \ket{\Psi(\bR,\bP)}$, then $\mathcal{T}\ket{\Psi(\bR,\bP)}$ is an eigenvector of $\hH_{PS}(\bR,-\bP)$ and   $E_{PS}(\bR,-\bP)  = E_{PS}(\bR,\bP)$. At the same time, however, note also  (ii) that as far as the electrons are concerned }  PSEST solves the electronic Schrodinger equation in the frame of moving nuclei in a fashion that violates time reversibility and Kramers' theorem.  \joeb{In other words, $\mathcal{T}\ket{\Psi(\bR,\bP)}$ need not be an eigenvector of $\hH_{PS}(\bR,\bP)$. Thus,}   PSEST can split apart two electronic states that are doubly degenerate within BO theory (as a function of $\bR$) -- but whose degeneracy is lifted as a function of $\bP$.   The splitting of these levels maps directly to the lambda doubling phenomena\cite{linqing:2026:lambda_doubling} and spin-rotation coupling\cite{linqing:2026:spin_rotation} that have been measured for small molecules very precisely.  This state of affairs is laid out visually in Fig. \ref{fig:kramers}. As shown in Refs. \citenum{linqing:2026:lambda_doubling, linqing:2026:spin_rotation} lambda doubling splittings and spin-rotation couplings can be extracted from the energy gaps between lower and  upper phase space electronic structure surfaces.

While the tests    described above were for small molecules, VCD and ROA signals can be extracted for  bulk materials\cite{vuilleumier:vcd:solidstate:2021,satoh:roa:solid:2026:prl}.  Moreover, because the difference between BO and PSEST is a small term proportional to an inverse nuclear mass, one would expect that PSEST and BO will differ most for systems with  small energy gaps and a great deal of electronic momenta -- for instance, metals and or low-gap extended systems.  Thus, we would anticipate that the future of PSEST will in fact involve more calculations with extended solids rather than for small molecules. Interestingly, in the context of Ehrenfest dynamics for periodic systems, Stengel\cite{stengel:traveling_pseudopoentials:prl:2026} and Mauri\cite{mauri:2026:traveling_pseudopotentials} and co-workers have recently proposed a method to boost pseudopotentials (akin to ETFs for pseudopotentials) so as to achieve angular and linear momentum conservation.  Another large potential application will be superconductivity\cite{tinkham_superconductivity}. By now, it has been well established that  superconductivity routinely involves  electron-phonon coupling, that $T_c$ changes with mass isotope, and superconductivity requires a  correct description of electronic momentum and super currents.   One provocative prediction is that PSEST will offer a new route to describe such effects. We note that  Eq. \ref{eq:HPSG1} involves a new two-electron term proportional to $\hbGamma  \cdot \hbGamma/(2M)$. This term is proportional to the inverse nuclear mass and was incorrectly identified as a one-electron term in Ref. \citenum{xuezhi:cpr:review:2026}. By changing the electron-electron interaction, one must wonder if PSEST will predict very new and rich physics.

We are not going to hypothesize any further about future experimental possibilities. For a more thorough discussion of these topics, including chiral induced spin selectivity (CISS)\cite{waldeck:2024:chemrev:ciss}, see Ref. \citenum{xuezhi:cpr:review:2026}.

\subsection{The Optimal $\Gamma$ and Approximations Thereof}

From a theoretical point of view, one of the most important questions going forward  is the question of how best to choose the $\hbGamma$ operator.  On the one hand, one could stick with the form given above  in Eq. \ref{eq:newgamma'} for $\hbGamma'$ (and 
more generally Eqs. 41-45 of Ref. \citenum{xuezhi:cpr:review:2026} for $\hbGamma''$, etc.).  On the other hand, one could imagine developing an   entirely new form for $\hbGamma$.  For the former case, one must also ask: how should we best pick the parameter $\sigma$ in Eq. \ref{eq:theta}?  We note that we have had some luck recently using the Becke weights\cite{becke:1988:jcp:beckegrid,frisch:cpl:1996:beckegrid,toddmartinez:book:beckegrid,curlee:becke:2026}  for $\theta$ instead of the Hirshfeld weights\cite{hirshfeld:1977:weights}  in Eq. \ref{eq:theta}.  Finally, as discussed in Sec. \ref{sec:curl} above, 
in 1D, one can develop a 
curl-free $\hbGamma$ but no such development appears possible in 3D. That being said, can we build $\hbGamma$ operators with explicit curl-free and curl-containing components, which might allow for us to capture as many PSEST effects as possible in a robust fashion in the future?

There are indeed many open questions about choosing the optimal form of $\hbGamma.$  Lastly, in the spirit of DFT, one could also imagine developing a $\hbGamma$ operator that itself depends on the electronic density, i.e. let $\hbGamma$ become an electronic-nuclear exchange correlation functional. 
In fact, suppose we diagonalize $\left( \frac{(\bP - i \hbar \hbGamma)^2}{2M} + V \right) \ket{\psi_0}  = E_0 \ket{\psi_0}$, and set $\hrho^{PS}_0 =\ket{\psi_0}\bra{\psi_0}$;  
one might  wonder if there is a $\hbGamma$ operator for which $W^{-1} \hrho^{PS}_0= \hrho^{exact}_0$ where
$\hrho_{exact}(\br',\bR',\br,\bR) = \Psi^{exact}_0(\br',\bR')\Psi^{exact}(\br,\bR)$
and $\Psi^{exact}_0(\br,\bR)$ is the exact ground state.
Many questions remain, and more interactions with mathematicians and physicists will be needed to give us guidance as to how best to construct and optimize $\hbGamma$.





\subsection{The Question of Spin $\Gamma'''$}

One critical, and still somewhat open question,  for future research in PSEST centers on the question of how to treat the electronic spin.  We began this account discussing the failures of surface hopping to account for electronic momentum and how that failure led us to  phase space electronic structure theory.   Of course, when discussing electronic states of matter, electronic spin is a crucial component of the angular momentum  and one of the critical goals of PSEST is to ensure that dynamics recover the correct spin properties. 

When performing a phase space electronic structure calculation and diagonalizing Eq. \ref{eq:diagHShenvi},
one unavoidable question is whether or not we should include $\hbGamma'''$ (see Eq. \ref{gammappp}).  Physically $\hbGamma'$ corresponds to running electronic structure calculations in a frame that is linearly boosted  relative to a local center of mass and  $\hbGamma''$ corresponds running electronic structure calculations where electronic motion is relative to a non-inertial body  frame (which produces a coriolis force).  $\hbGamma'''$ corresponds to running electronic structure calculations in a frame whereby the electronic spin (like the orbitals) is also solved in the body frame associated with the local nuclei (which leads to a spin-coriolis force).

One must then wonder: Does including $\hbGamma'''$ lead to improved potential energy surfaces or not? The answer to this question must depend on the size of the spin orbit coupling.  Imagine  a molecule which is stationary at time zero and whose spin is in direction $\vec{s}$ in the lab frame.  If one now gives the molecule a push so as to initiate a rotation, does the spin rotate with the molecule or does the spin keep its direction in the lab frame? The result depends of course on the magnitude of the  spin-orbit coupling. With zero or a very weak SOC, the spin direction must stay fixed in the lab frame. With strong SOC, the spin direction must change and will be influenced by the motion of the nuclei.  Thus, the decision as to whether or not to include $\hbGamma'''$ must require information about the magnitude of the SOC and the time scale for for the rotational motion coupled to the spin, from which we may surmise whether the spin should  remain constant in the lab frame or  the body frame. 
Working out exactly the cutoffs for SOC and the conditions for which $\hbGamma'''$  should be included will be essential if we want to accurately capture the entanglement of nuclear electronic and spin degrees of freedom.

\subsection{Missed Opportunities and Connection to Other Theories}
In closing, we believe the realm of phase space electronic structure is exploding in the present time, as the realization sinks in that one can realistically go beyond the BO framework and explore questions of electronic structure without freezing the nuclei. Upon reflection, one might wonder why the idea of generating electronic states that depended on $\bR$ and $\bP$ did not come sooner. The idea is simple and there were indeed many opportunities. Notably, our developments in this field were not (but probably should have been!) guided by previous work in electronic friction\cite{suhl:1975:prb_elfriction,tully:1995:electronic_friction} and dynamics at metal surfaces, where it is known that molecules feel a drag and random force -- and a pseudomagnetic field when out of equilibrium\cite{vonOppen:2012:elfriction_belstein,hedegard:2010:nano_berry}. At the time that our research groups worked in this area\cite{dou:2017:prl,dou:2018:prb_vonOppen_noneq}, we did not consider the implications of Berry forces as far as momentum conservation is concerned; that realization came  only later\cite{xuezhi:2023:total_ang_bomd} when we began to work with SOC explicitly\cite{teh:equilibrium:berryforce:prb,teh:non_equilibrium:berryforce:prb}.

It is also worth noting that our progress in PSEST was  not guided by knowledge of small molecule chemistry, though the connection is now clear.  After all, within the realm of high resolution spectroscopy,   experimentalists have measured rovibronic energies for decades and have long  recognized that, in order to theory to match experiment, they needed to work in internal coordinates,
 transforming away the molecular translations (by working relative to the total center of mass) and isolating the effects of rotations of the body frame\cite{bunker:book:2006}.  Through the lens of PSEST, we now understand that such transforming away of the translations and taming of the rotations is equivalent to incorporating some components of the derivative coupling back into the electronic hamiltonian.  In other words, PSEST generalizes the molecular transformations previously used in high-resolution spectroscopy, but now including local vibrations and internal torsions, effectively treating every nuclear center as a local center of mass for relative electronic motion. As such, while PSEST can certainly be applied to small molecule chemistry (and effectively was applied years ago by high resolution spectroscopists), we now realize that PSEST can also be applied in the condensed phase, e.g.  to solutions of floppy molecules and/or crystals. 
 In this vein, one must presume the  development of PSEST was severely slowed down by the disconnect between electronic structure theorists and gas phase spectroscopists.


One can hypothesize many other theories as well for why PSEST was not pursued earlier pertains to the question of exactness. Perhaps the answer lies in the discomfort that chemists and physicists inevitably feel in working with Wigner-Weyl transforms and symbols vs operators, and the fact that 
extracting the exact answer is easier with the BO picture than with PS (or they were scared by Heller's warnings about convergence\cite{heller:1976:wigner_problems})? 
Or perhaps  gas spectroscopists insisted on having exact, unique $\hbGamma$ operators and were uncomfortable with the non-unique $\theta$ partitions of unity that we have been forced to introduce for molecules that can both rotate {\em and} vibrate. 
Or perhaps the answer is that most of the novel effects captured by PSEST (relative to BO) are small -- because of the mass difference between electrons and nuclei -- and there was no focus on the exotic chemistry that one finds  sometimes with spins and magnetic fields; only time will tell how many  novel applications we can find.   Again, there are many plausible explanations and 
there were many missed opportunities: as this perspective should make clear,
our  own developments in this field came as the last result after our  own failures with surface hopping dynamics -- and filled with many missteps along the way.

Looking forward, however, a few other research groups are now moving in the same direction as are we. Most notably, the research groups of Polkovnikov and Chandran  have begun to transition from more traditional models of quantum geometry\cite{polkovnikov:2017:quantum_geometry} to a PSEST approach--which they call a moving Born-Oppenheimer approximation [MBOA] and which resembles Shenvi's approach more than ours\cite{polkovnikov:2026:pnas}.   As such,  there is hope that PSEST will slowly attract more attention in both the chemistry and physics communities. Moreover, as discussed above, the question of how to apply PSEST in the context of solid state electronic structure and interfacial chemistry remains a critical area for future development. Finally, the fact that recent research into the exact factorization approach\cite{gross:2022:prl:angmom} has focused on nuclear rotations and angular momentum conservation in electronic media  would further seem to suggest that still another group of dynamicists and electronic structure theorists are also coming to the same realizations about the limits of BO theory--although PSEST is very different from exact factorization theory. After all, exact factorization aims to solve the entire problem of nuclear-electronic correlation in one shot, while PSEST has the much more modest (but we would arguable tractable) goal of  replacing $E(\bR)$ with an $E(\bR,\bP)$.

\section{Conclusions and Outlook}

It is our hope that the presentation above has offered the reader both a physical as well as a historical overview of the big questions behind phase-space electronic structure theory.  The approach captures many subtle problems in non-adiabatic theory that are not easily solved  through  other means.   Within our experience, the theory  originated  from a combination of frustration with Tully's surface hopping algorithm's ability to treat SOC and the subsequent realization that Shenvi's phase space surface hopping protocol might hold the solution.  This motivation, combined with  our desire for {\em ab initio} electronic structure capable of treating degenerate states with spin degrees that conserved total momentum, leads to Eq. \ref{eq:HPSG1}. 

For the overly optimistic reader, it is critical to emphasize that phase space electronic structure does not solve the electronic structure problem; one must still deal with electron correlation that is produced by the two-electron Coulomb potential. PSEST also does not solve the problem of curve crossings or conical intersections or the need for surface hopping dynamics. Although one would presume that conical intersections are more rare in PSEST than within BO theory (because PSEST CIs have codimension 3\cite{titouan:2025:ci_phase_space:jpcl} while BO CIs have codimension 2),   PSEST certainly cannot eliminate such crossings altogether.  Sharp derivative couplings will still arise near a strong curve crossing/intersection where electronic character changes, and in such case, a dynamical approach like surface hopping, now  phase space surface hopping\cite{shenvi:2009:jcp_pssh}, will be required.  That being said, from a nonadiabatic point of view, the strength of  PSEST  is that one  can treat the small derivative couplings that arise far from a curve crossing (e.g. during  rotational or librational motion) and  eliminate many of the small derivative couplings that always arise everywhere in space. Most poignantly, unlike the case of BO FSSH, phase space surface dynamics will effectively not hop when the nuclei are executing  translational or rotational motion (where the energy gap does not change)\cite{bian:2024:pssh_translations_rotations}.

\begin{figure}
    \centering
    \includegraphics[width=0.9\linewidth]{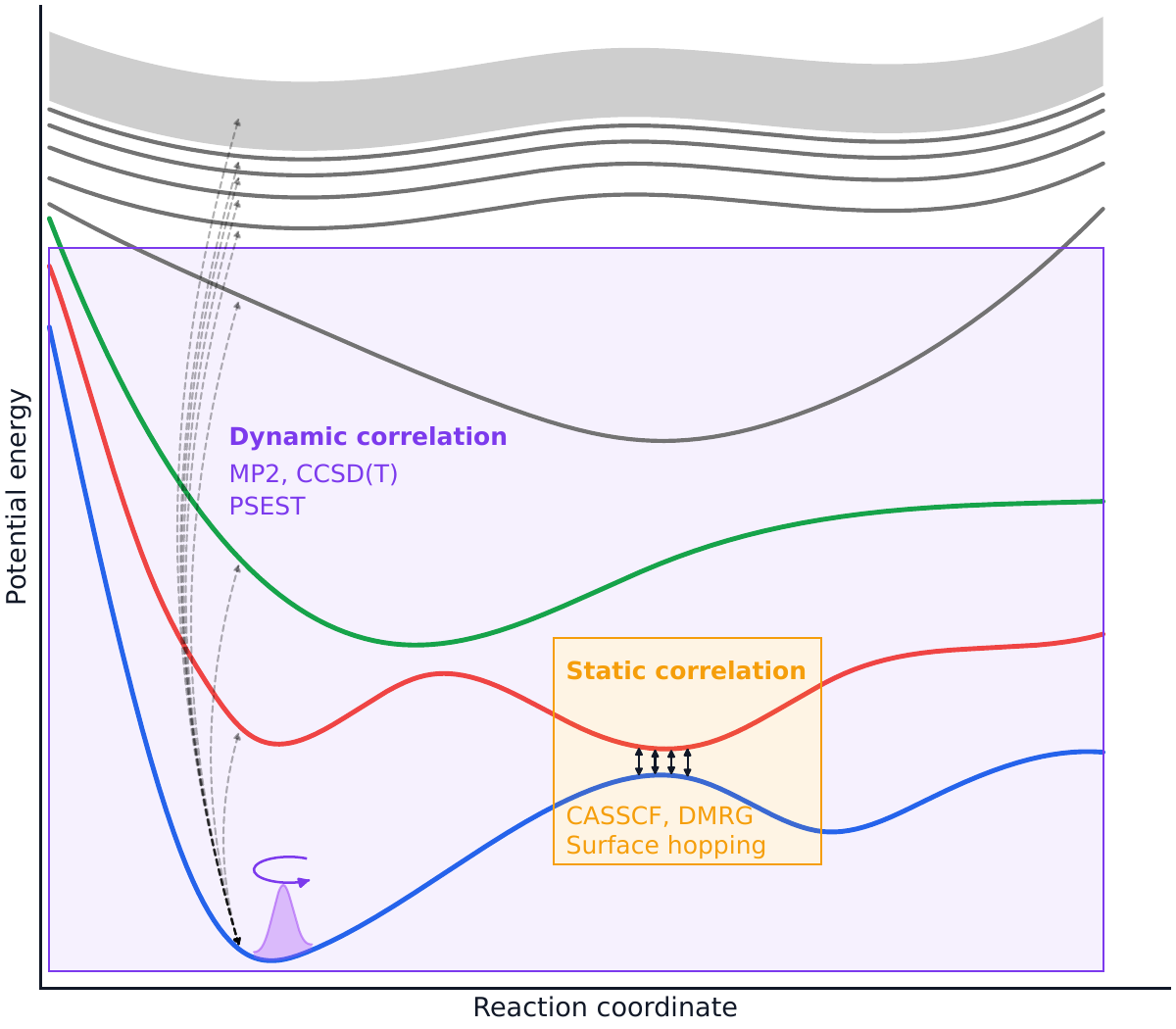}
\caption{An intuitive picture of how to understand correlation within the confines of  PSEST.  On the one hand, {\em static  correlation} occurs near curve crossings between two (or a handful of) strongly coupled electronic states. In the context of electron-electron correlation, static correlation describes multireference effects, which are usually solved with CASSCF or DMRG methods.  In the context of electron-nuclear coupling, static correlation occurs when one does not know which surface to move along and one often jumps or hops between surfaces.  On the other hand, {\em dynamic correlation} occurs everywhere in configuration space and results from the weak interactions  between any one electronic state and the larger manifold of of all other excited electronic states. In the context of electron-electron correlation, dynamic correlation is often equated to electronic scattering, which is usually treated perturbatively with MP2 or non-perturbatively with coupled-cluster theory (e.g. CCSD(T)) methods -- but always based on single references.  In the context of electron-nuclear correlation, dynamic correlation involves electronic coriolis forces, electronic centrifugal forces, and the range of other forces that arise when the electrons are perturbed by accelerating nuclei. Mathematically, these correlations arise when a given electronic state interacts weakly with many other electronic states through nuclear momentum induced coupling; such
couplings are never exactly zero\cite{mbaer:1975:cpl,vv:2012:chemrev,littlejohn:2022:jcp:parallel}, even if the gap between states is many electron volts. In fact, these couplings can vanish as slowly as the inverse of the energy gap. 
    One incorporates these effects  by creating phase space electronic surfaces $E_j(\bR,\bP)$. }
    \label{fig:cpr}
\end{figure}

In closing, we have found the schematic diagram in Figure \ref{fig:cpr} to be very helpful.  Far from curve crossings, phase space electronic structure theory allows us to conserve the total nuclear plus electronic momentum by capturing the interaction of one active state interacting weakly with  many higher lying electronic states through a momentum-dependent coupling.  Near a curve crossing, however, one must still hop between two strongly, nearly degenerate coupled surfaces if one wants to capture electronic relaxation correctly; one can also construct phase space diabatic states\cite{zain:shin_metiu_diabatization:2026_first_paper}. We refer to the processes above as "dynamic" and "static" correlation (which draws on the language used in electron-electron correlation theory\cite{reiher:2016:static_dynamics_correlation}). 
Admittedly, this framework challenges  the separation between thermodynamics and kinetics that we are comfortable with and engaging with PSEST surfaces will necessarily require us to develop a great deal more intuition about coupled nuclear-electronic motion. That being said, we strongly feel that developing the necessary electronic structure codes and cultivating a new intuition is  worth the cost, if only given the large number of potential applications in spin chemistry and magnetic chemistry.
In writing this perspective, with a focus on retracing the origins of PSEST (in combination with the more technical review in Ref. \citenum{xuezhi:cpr:review:2026}),
our hope is that we will attract more young theoretical chemists and physicists to this bourgeoning discipline.

\section{Acknowledgments}
This perspective  was supported by the U.S. Air Force Office of Scientific Research (USAFOSR)
under Grant No. FA9550-23-1-0368 and the National Science Foundation
under Grant No. CHE-2422858.  We acknowledge the DoD High Performance Computing Modernization Program for computer time. GPT-5.4 by OpenAI was used on March 27, 2026 and April 3, 2026 through ChatGPT to render part of Fig. \ref{fig:linqing}. 

\section{Data Availability Statement}
Data sharing is not applicable to this article as no new data were created or analyzed in this study.

\bibliography{finalbib}

@string{JCP = "{\em J. Chem. Phys.}"}

@string{JPCA = "{\em J. Phys. Chem. A}"}

@string{JPCB = "{\em J. Phys. Chem. B}"}

@string{JPCL = "{\em J. Phys. Chem. Lett.}"}

@string{ARPC = "{\em Ann. Rev. Phys. Chem. }"}

@string{CR = "{\em Chem. Rev.}"}

@string{RMP = "{\em Rev. of Mod. Phys.}"}

@string{PRL= "{\em Phys. Rev. Lett.}"}

@string{PRX= "{\em Phys. Rev. X}"}

@string{PR = "{\em Phys. Rev. }"}

@string{PRA= "{\em Phys. Rev. A}"}

@string{PRB= "{\em Phys. Rev. B}"}

@string{CPL = "{\em Chem. Phys. Lett.}"}

@string{CP = "{\em Chem. Phys. }"}

@string{JCTC  = "{\em J. Chem. Theory Comp.}"}

@string{ACP = "{\em Adv. Chem. Phys.}"}

@string{IJQC = "{\em Int. J. Quant. Chem.}"}

@string{MP = "{\em Mol. Phys.}"}

@string{PRSLA = "{\em Proc. Roy. Sci. Lond. A }"}

@string{JCP = "{ J. Chem. Phys.}"}

@string{JPCA = "{ J. Phys. Chem. A}"}

@string{JPCB = "{ J. Phys. Chem. B}"}

@string{JPCL = "{ J. Phys. Chem. Lett.}"}

@string{ARPC = "{ Ann. Rev. Phys. Chem. }"}

@string{CR = "{ Chem. Rev.}"}

@string{RMP = "{ Rev. of Mod. Phys.}"}

@string{PRL= "{ Phys. Rev. Lett.}"}

@string{PRL= "{ Phys. Rev. X}"}

@string{PR = "{ Phys. Rev. }"}

@string{PRA= "{ Phys. Rev. A}"}

@string{PRB= "{ Phys. Rev. B}"}

@string{CPL = "{ Chem. Phys. Lett.}"}

@string{CP = "{ Chem. Phys. }"}

@string{JCTC  = "{ J. Chem. Theory Comp.}"}

@string{ACP = "{ Adv. Chem. Phys.}"}

@string{IJQC = "{ Int. J. Quant. Chem.}"}

@string{MP = "{ Mol. Phys.}"}

@string{PRSLA = "{ Proc. Roy. Sci. Lond. A }"}

@string{JCP = "Journal of Chemical Physics"}

@string{JPCA = "Journal of Physical Chemistry A"}

@string{JPCB = "Journal of Physical Chemistry B"}

@string{JPCL = "Journal of Physical Chemistry Letters"}

@string{ARPC = "Annual Reviews in Physical Chemistry"}

@string{CR = "Chemical Reviews"}

@string{RMP = "Reviews of Modern Physics"}

@string{PRL = "Physical Review Letters"}

@string{PRX = "Physical Review X"}

@string{PR = "Physical Review"}

@string{PRA = "Physical Review A"}

@string{PRB = "Physical Review B"}

@string{CPL = "Chemical Physics Letters"}

@string{CP = "Chemical Physics"}

@string{JCTC  = "Journal of Chemical Theory and Computation"}

@string{ACP = "Advances in Chemical Physics"}

@string{IJQC = "International Journal of Quantum Chemistry"}

@string{MP = "Molecular Physics"}

@string{PRSLA = "Proceedings of the Royal Society of London A "}

@book{
   szabo:ostlund,
   author = {A. Szabo and N. Ostlund},
   title = {Modern Quantum Chemistry: Introduction to Advanced Electronic Structure Theory},
   publisher = {Dover},
   address = {New Jersey},
   year = {1996}
}

@article{
   tully:2000:review,
   author = {J. C. Tully},
   title = {Chemical Dynamics at Metal Surfaces},
   journal=ARPC,
   volume = {51},
   number = {},
   pages = {153-178},
   year = {2000}
}

@book{
   jensen,
   author = {F. Jensen},
   title = {Introduction to Computational Chemistry},
   publisher = {Wiley},
   address = {England},
   year = {1999}
}

@book{
   nitzanbook,
   author = {A. Nitzan},
   title = {Chemical Dynamics in Condensed Phases},
   publisher = {Oxford University Press},
   address = {USA},
   year = {2006}
}

@article{
   mbaer:1975:cpl,
   author = {M. Baer},
   title = {Adiabatic and diabatic representations for atom-molecule collisions: Treatment of the collinear arrangement  },
   journal=CPL,
   volume = {35},
   number = {},
   pages = {112},
   year = {1975}
}

@article{
   tully:fssh,
   author = {J. C. Tully},
   title = {Molecular dynamics with electronic transitions},
   journal=JCP,
   volume = {93},
   number = {},
   pages = {1061-1071},
   year = {1990}
}

@article{
   tully:faraday:fssh,
   author = {J. C. Tully},
   title = {Mixed quantum-classical dynamics },
   journal={Faraday Discussions},
   volume = {110},
   number = {},
   pages = {407-419},
   year = {1998}
}

@article{
   kapral:1999:jcp,
   author = {R. Kapral and G. Ciccotti},
   title = {Mixed quantum-classical dynamics},
   journal=JCP,
   volume = {110},
   number = {},
   pages = {8919-8929},
   year = {1999}
}

@article{
   miller:2007:ananth,
   author = {N. Ananth and C. Venkataraman and W. H. Miller},
   title = {Semiclassical description of electronically nonadiabatic dynamics 
via the initial value representation},
   journal=JCP,
   volume = {127},
   number = {},
   pages = {084114},
   year = {2007}
}

@article{
   marcus:1956,
   author = {R. A. Marcus},
   title = {On the Theory of Oxidation-Reduction Reactions Involving Electron Transfer. I},
   journal=JCP,
   volume = {24},
   number = {},
   pages = {966-978},
   year = {1956}
}

@article{
   tretiak:2011:conj,
   author = { T. Nelson and S. Fernandez-Alberti and V. Chernyak and A. E. Roitberg and S. Tretiak },
   title = { Nonadiabatic Excited-State Molecular Dynamics Modeling of Photoinduced Dynamics in Conjugated Molecules }, 
   journal=JPCB,
   volume = {115},
   number = {},
   pages = {5402-5414},
   year = {2011}
}

@article{
   yarkony:1992:advchemphys,
   author = {B. H. Lengsfield and D. R. Yarkony},
   title = {Nonadiabatic interactions between potential energy surfaces: Theory and applications},
   journal=ACP,
   volume = {82 (part 2)},
   number = {},
   pages = {1-71},
   year = {1992}
}

@article{
   yarkony:1984:jcp_dercouple,
   author = {B. H. Lengsfield and  P. Saxe and D. R. Yarkony},
   title = {On the evaluation of nonadiabatic coupling matrix elements using SA‐MCSCF/CI wave functions and analytic gradient methods. I},
   journal=JCP,
   volume = {81},
   number = {},
   pages = {4549},
   year = {1984}
}

@article{
   fatehi:2011:dercouple,
   author = {S. Fatehi and E. Alguire and Y. Shao and J. E. Subotnik},
   title = {Analytical Derivative Couplings Between Configuration Interaction Singles States with Built-in Translation Factors for Translational Invariance },
   journal=JCP,
   volume = {135},
   number = {},
   pages = {234105},
   year = {2011},
   doi = {10.1063/1.3665031}
}

@article{
   fatehi:2012:dercouple,
   author = {S. Fatehi and J. E. Subotnik},
   title = { Derivative Couplings with Built-In Electron-Translation Factors: Application to Benzene}, 
   journal=JPCL,
   volume = {3},
   number = {},
   pages = {2039-2043},
   year = {2012}
}

@article{
   subotnik:2013:qcle_fssh_derive,
   author = {J. E. Subotnik and W. Ouyang and B. R. Landry},
   title = {Can we derive Tully's surface-hopping algorithm from the semiclassical quantum Liouville equation: Almost, but only with decoherence},
   journal=JCP,
   volume = {139},
   number = {},
   pages = {214107},
   year = {2013}
}

@article{
   landry:2014:abinitio_closs,
   author = {B. R. Landry and J. E. Subotnik},
   title = {Quantifying the lifetime of triplet energy transfer processes in organic chromophores: A case study of 4-(2-naphthylmethyl)benzaldehyde},
   journal=JCTC,
   volume = {10},
   number = {},
   pages = {4253-4263},
   year = {2014}
}

@article{
   coker:2012:iterative,
   author = {P. Huo and D. Coker},
   title = { Consistent schemes for non-adiabatic dynamics derived from partial linearized density matrix propagation },
   journal=JCP,
   volume = {137},
   number = {},
   pages = {22A535},
   year = {2012}
}

@article{
   coker:2008:iterative,
   author = {E. R. Dunkel and S. Bonella and D. Coker},
   title = {Iterative linearized approach to nonadiabatic dynamics},
   journal=JCP,
   volume = {129},
   number = {},
   pages = {114106},
   year = {2008}
}

@article{
   tully:1995:electronic_friction,
   author = {M. Head-Gordon and J. C. Tully},
   title = {Molecular dynamics with electronic frictions},
   journal=JCP,
   volume = {103},
   number = {},
   pages = {10137-10145},
   year = {1995}
}

@article{
   shs:1994:protons,
   author = {Sharon Hammes-Schiffer and J.C. Tully},
   title = "Proton transfer in solution: Molecular dynamics with quantum transitions",
   journal = JCP,
   year = "1994",
   volume = "101",
   number = "6",
   pages = "4657-4667",
}

@article{
   kapral:2008:jcp_pbme,
   author = { H. Kim and A. Nassimi and R. Kapral},
   title = {Quantum-classical liouville dynamics in the mapping basis},
   journal=JCP,
   volume = {129},
   number = {},
   pages = {084102},
   year = {2008}
}

@ARTICLE{
  hynes:ci_fssh:review,
 AUTHOR={Malhado, João Pedro  and  Bearpark, Michael John  and  Hynes, James T},
TITLE={Non-adiabatic dynamics close to conical intersections and the surface hopping perspective},
JOURNAL={Frontiers in Chemistry},
VOLUME={2},
YEAR={2014},
NUMBER={97}
}

@article{
miller:2001:jpcareview,
author = {Miller, William H.},
title = {The Semiclassical Initial Value Representation:  A Potentially Practical Way for Adding Quantum Effects to Classical Molecular Dynamics Simulations},
journal = JPCA,
volume = {105},
number = {13},
pages = {2942-2955},
year = {2001},
}

@article{
   miller:2012:coherence,
   author = "Miller, William H.",
   title = "Perspective: Quantum or classical coherence?",
   journal = JCP,
   year = "2012",
   volume = "136",
   number = "21",
   pages = 210901,
}

@article {
barbatti:2011:review,
author = {Barbatti, Mario},
title = {Nonadiabatic dynamics with trajectory surface hopping method},
journal = {Wiley Interdisciplinary Reviews: Computational Molecular Science},
volume = {1},
number = {4},
publisher = {John Wiley & Sons, Inc.},
issn = {1759-0884},
pages = {620--633},
year = {2011},
}

@book{
   tannor:quantumbook,
   author = {D. Tannor},
   title = {Introduction to Quantum Mechanics: A Time-Dependent Perspective},
   publisher = {University Science Books},
   address = {},
   year = {2006}
}

@article{
jain:2016:fast,
author = {Jain, Amber and Alguire, Ethan and Subotnik, Joseph E.},
title = {An Efficient, Augmented Surface Hopping Algorithm That Includes Decoherence for Use in Large-Scale Simulations},
journal = JCTC,
volume = {12},
number = {11},
pages = {5256-5268},
year = {2016},
doi = {10.1021/acs.jctc.6b00673},
    note ={PMID: 27715036},

URL = {
        http://dx.doi.org/10.1021/acs.jctc.6b00673

},
eprint = {
        http://dx.doi.org/10.1021/acs.jctc.6b00673

}

}

@article{
qchem4,
author = {Yihan Shao and Zhengting Gan and Evgeny Epifanovsky and Andrew T.B. Gilbert and Michael Wormit and Joerg Kussmann and Adrian W. Lange and Andrew Behn and Jia Deng and Xintian Feng and Debashree Ghosh and Matthew Goldey and Paul R. Horn and Leif D. Jacobson and Ilya Kaliman and Rustam Z. Khaliullin and Tomasz Kuś and Arie Landau and Jie Liu and Emil I. Proynov and Young Min Rhee and Ryan M. Richard and Mary A. Rohrdanz and Ryan P. Steele and Eric J. Sundstrom and H. Lee Woodcock III and Paul M. Zimmerman and Dmitry Zuev and Ben Albrecht and Ethan Alguire and Brian Austin and Gregory J. O. Beran and Yves A. Bernard and Eric Berquist and Kai Brandhorst and Ksenia B. Bravaya and Shawn T. Brown and David Casanova and Chun-Min Chang and Yunqing Chen and Siu Hung Chien and Kristina D. Closser and Deborah L. Crittenden and Michael Diedenhofen and Robert A. DiStasio Jr. and Hainam Do and Anthony D. Dutoi and Richard G. Edgar and Shervin Fatehi and Laszlo Fusti-Molnar and An Ghysels and Anna Golubeva-Zadorozhnaya and Joseph Gomes and Magnus W.D. Hanson-Heine and Philipp H.P. Harbach and Andreas W. Hauser and Edward G. Hohenstein and Zachary C. Holden and Thomas-C. Jagau and Hyunjun Ji and Benjamin Kaduk and Kirill Khistyaev and Jaehoon Kim and Jihan Kim and Rollin A. King and Phil Klunzinger and Dmytro Kosenkov and Tim Kowalczyk and Caroline M. Krauter and Ka Un Lao and Adèle D. Laurent and Keith V. Lawler and Sergey V. Levchenko and Ching Yeh Lin and Fenglai Liu and Ester Livshits and Rohini C. Lochan and Arne Luenser and Prashant Manohar and Samuel F. Manzer and Shan-Ping Mao and Narbe Mardirossian and Aleksandr V. Marenich and Simon A. Maurer and Nicholas J. Mayhall and Eric Neuscamman and C. Melania Oana and Roberto Olivares-Amaya and Darragh P. O’Neill and John A. Parkhill and Trilisa M. Perrine and Roberto Peverati and Alexander Prociuk and Dirk R. Rehn and Edina Rosta and Nicholas J. Russ and Shaama M. Sharada and Sandeep Sharma and David W. Small and Alexander Sodt and Tamar Stein and David Stück and Yu-Chuan Su and Alex J.W. Thom and Takashi Tsuchimochi and Vitalii Vanovschi and Leslie Vogt and Oleg Vydrov and Tao Wang and Mark A. Watson and Jan Wenzel and Alec White and Christopher F. Williams and Jun Yang and Sina Yeganeh and Shane R. Yost and Zhi-Qiang You and Igor Ying Zhang and Xing Zhang and Yan Zhao and Bernard R. Brooks and Garnet K.L. Chan and Daniel M. Chipman and Christopher J. Cramer and William A. Goddard III and Mark S. Gordon and Warren J. Hehre and Andreas Klamt and Henry F. Schaefer III and Michael W. Schmidt and C. David Sherrill and Donald G. Truhlar and Arieh Warshel and Xin Xu and Alán Aspuru-Guzik and Roi Baer and Alexis T. Bell and Nicholas A. Besley and Jeng-Da Chai and Andreas Dreuw and Barry D. Dunietz and Thomas R. Furlani and Steven R. Gwaltney and Chao-Ping Hsu and Yousung Jung and Jing Kong and Daniel S. Lambrecht and WanZhen Liang and Christian Ochsenfeld and Vitaly A. Rassolov and Lyudmila V. Slipchenko and Joseph E. Subotnik and Troy Van Voorhis and John M. Herbert and Anna I. Krylov and Peter M.W. Gill and Martin Head-Gordon},
title = {Advances in molecular quantum chemistry contained in the Q-Chem 4 program package},
journal =  MP,
volume = {113},
number = {2},
pages = {184-215},
year = {2015}
}

@article{
levine:2014:trivial,
author = {Meek, Garrett A. and Levine, Benjamin G.},
title = {Evaluation of the Time-Derivative Coupling for Accurate Electronic State Transition Probabilities from Numerical Simulations},
journal = JPCL,
volume = {5},
number = {13},
pages = {2351-2356},
year = {2014},
}

@article{
gonzalez:2011:sharc,
author = {Richter, Martin and Marquetand, Philipp and González-Vázquez, Jesús and Sola, Ignacio and González, Leticia},
title = {SHARC: ab Initio Molecular Dynamics with Surface Hopping in the Adiabatic Representation Including Arbitrary Couplings},
journal = JCTC,
volume = {7},
number = {5},
pages = {1253-1258},
year = {2011},
}

@article{
vonOppen:2012:elfriction_belstein,
author="Niels Bode and Silvia Viola Kusminskiy and Reinhold Egger and Felix von Oppen",
title="Current-induced forces in mesoscopic systems: A scattering-matrix approach",
journal="Beilstein Journal of Nanotechnology",
year="2012",
volume="3",
pages="144-162",
}

@article{
   martinez:1996:jcp_nai,
   author = "Martinez, Todd J. and Levine, R. D.",
   title = "First‐principles molecular dynamics on multiple electronic states: A case study of NaI",
   journal = JCP,
   year = "1996",
   volume = "105",
   number = "15", 
   pages = "6334-6341",
}

@article{
    lischka:2012:jcp_localized_diabatization,
   author = "Plasser, Felix and Granucci, Giovanni and Pittner, Jiri and Barbatti, Mario and Persico, Maurizio and Lischka, Hans",
   title = "Surface hopping dynamics using a locally diabatic formalism: Charge transfer in the ethylene dimer cation and excited state dynamics in the 2-pyridone dimer",
   journal = JCP,
   year = "2012",
   volume = "137",
   number = "22", 
   pages = "22A514",
}

@article{
   shenvi:2009:jcp_pssh,
   author = "Shenvi, Neil",
   title = "Phase-space surface hopping: Nonadiabatic dynamics in a superadiabatic basis",
   journal = jcp,
   year = "2009",
   volume = "130",
   number = "12",
   pages = 124117,
}

@article{
   berry:1987:superadiabat,
   author = {M. V. Berry},
   title = {Quantum Phase Corrections from Adiabatic Iteration},
   journal=PRSLA,
   volume = {414},
   number = {},
   pages = {31-46},
   year = {1987}
}

@article {          
       berry:1984:berryphase,
        author = {Berry, M. V.},
        title = {Quantal Phase Factors Accompanying Adiabatic Changes},
        volume = {392},
        number = {1802},
        pages = {45--57},
        year = {1984},
        journal = PRSLA,
}

@article{
    izmaylov:2016:jpc_dboc_pssh,
   author = "Gherib, Rami and Ye, Liyuan and Ryabinkin, Ilya G. and Izmaylov, Artur F.",
   title = "On the inclusion of the diagonal Born-Oppenheimer correction in surface hopping methods",
   journal = JCP,
   year = "2016",
   volume = "144",
   number = "15",
   pages = 154103,
}

@article{
   subotnik:2016:arpc,
   author = {Joseph E. Subotnik and  Amber Jain and  Brian Landry and  Andrew Petit and  Wenjun Ouyang and Nicole Bellonzi},
   title = {Understanding the Surface Hopping View of Electronic Transitions and Decoherence},
   journal=ARPC,
   volume = {67},
   number = {},
   pages = {387-417},
   year = {2016}
}

@article{
vv:2012:chemrev,
author = {Kaduk, Benjamin and Kowalczyk, Tim and Van Voorhis, Troy},
title = {Constrained Density Functional Theory},
journal = CR,
volume = {112},
number = {1},
pages = {321-370},
year = {2012},
doi = {10.1021/cr200148b},
}

@article{
   suhl:1975:prb_elfriction,
  title = {Brownian motion model of the interactions between chemical species and metallic electrons: Bootstrap derivation and parameter evaluation},
  author = {d'Agliano, E. G. and Kumar, P. and Schaich, W. and Suhl, H.},
  journal = PRB,
  volume = {11},
  pages = {2122--2143},
  year = {1975},
  doi = {10.1103/PhysRevB.11.2122},
}

@article{
kapral:2016:chemphys_fssh,
title = "Surface hopping from the perspective of quantum–classical Liouville dynamics",
journal = CP,
volume = "481",
pages = "77 - 83",
year = "2016",
note = "Quantum Dynamics and Femtosecond Spectroscopy dedicated to Prof. Vladimir Y. Chernyak on the occasion of his 60th birthday",
issn = "0301-0104",
doi = "https://doi.org/10.1016/j.chemphys.2016.05.016",
url = "http://www.sciencedirect.com/science/article/pii/S0301010416302725",
author = "Raymond Kapral",
}

@article{
granucci:2012:fssh_spinorbit,
author = {Giovanni Granucci and Maurizio Persico and Gloria Spighi},
title = {Surface hopping trajectory simulations with spin-orbit and dynamical couplings},
journal = JCP,
volume = {137},
number = {22},
pages = {22A501},
year = {2012},
doi = {10.1063/1.4707737},
URL = { 
        http://dx.doi.org/10.1063/1.4707737
},
}

@article{
  dou:2017:prl, 
  title = {Born-Oppenheimer Dynamics, Electronic Friction, and the Inclusion of Electron-Electron Interactions},
  author = {Dou, Wenjie and Miao, Gaohan and Subotnik, Joseph E.},
  journal = PRL,
  volume = {119},
  issue = {4},
  pages = {046001},
  numpages = {6},
  year = {2017},
  url = {https://link.aps.org/doi/10.1103/PhysRevLett.119.046001}
}

@article{
  pechukas:1969:pr,
  title = {Time-Dependent Semiclassical Scattering Theory. II. Atomic Collisions},
  author = {Pechukas, Philip},
  journal = PR,
  volume = {181},
  issue = {1},
  pages = {174--185},
  numpages = {0},
  year = {1969},
  month = {May},
  publisher = {American Physical Society},
  doi = {10.1103/PhysRev.181.174},
  url = {https://link.aps.org/doi/10.1103/PhysRev.181.174}
}

@article{
mead:1979:noncrossing,
author = {Mead,C. Alden },
title = {The ’’noncrossing’’ rule for electronic potential energy surfaces: The role of time‐reversal invariance},
journal = JCP,
volume = {70},
number = {5},
pages = {2276-2283},
year = {1979},
doi = {10.1063/1.437733},

URL = { 
        https://doi.org/10.1063/1.437733
    
},
eprint = { 
        https://doi.org/10.1063/1.437733
    
}

}

@article{
meadtruhlar:1979:berry,
author = {Mead,C. Alden  and Truhlar,Donald G. },
title = {On the determination of Born–Oppenheimer nuclear motion wave functions including complications due to conical intersections and identical nuclei},
journal = JCP,
volume = {70},
number = {5},
pages = {2284-2296},
year = {1979},
doi = {10.1063/1.437734},

URL = { 
        https://doi.org/10.1063/1.437734
    
},
eprint = { 
        https://doi.org/10.1063/1.437734
    
}

}

@article{ 
  hedegard:2010:nano_berry,
  title={Blowing the fuse: Berry’s phase and runaway vibrations in molecular conductors},
  author={Lu, Jing-Tao and Brandbyge, Mads and Hedeg{\aa}rd, Per},
  journal={Nano letters},
  volume={10},
  number={5},
  pages={1657--1663},
  year={2010},
  publisher={ACS Publications}
}

@article{
miao:2019:fssh:complex,
author = {Miao,Gaohan  and Bellonzi,Nicole  and Subotnik,Joseph },
title = {An extension of the fewest switches surface hopping algorithm to complex Hamiltonians and photophysics in magnetic fields: Berry curvature and “magnetic” forces},
journal = JCP,
volume = {150},
number = {12},
pages = {124101},
year = {2019},
doi = {10.1063/1.5088770},

URL = { 
        https://doi.org/10.1063/1.5088770
    
},
eprint = { 
        https://doi.org/10.1063/1.5088770
    
}

}

@article{
berry:1993:royal:half_classical,
   author = {M. V. Berry and J. M. Robbins},
   title = {Chaotic classical and half-classical adiabatic reactions: geometric magnetism and deterministic friction},
   journal=PRSLA,
   volume = {442},
   number = {},
   pages = {659-672},
   year = {1993},
   doi= {https://doi.org/10.1098/rspa.1993.0127}
}

@article{
herman:1984:jcp:rescaling_direction,
author = {Herman,Michael F. },
title = {Nonadiabatic semiclassical scattering. I. Analysis of generalized surface hopping procedures},
journal = JCP,
volume = {81},
number = {2},
pages = {754-763},
year = {1984},
doi = {10.1063/1.447708},
URL = { 
        https://doi.org/10.1063/1.447708
},
eprint = { 
        https://doi.org/10.1063/1.447708
}
}

@article{
  bian:2021:perspective,
  title={Modeling nonadiabatic dynamics with degenerate electronic states, intersystem crossing, and spin separation: A key goal for chemical physics},
  author={Bian, Xuezhi and Wu, Yanze and Teh, Hung-Hsuan and Zhou, Zeyu and Chen, Hsing-Ta and Subotnik, Joseph E},
  journal=JCP,
  volume={154},
  number={11},
  pages={110901},
  year={2021},
  publisher={AIP Publishing LLC}
}

@article{
   bates:1958:etf,
   author = {D. R. Bates and R. McCarroll},
   title = {Electron capture in slow collisions},
   journal={Proc. R. Soc. A},
   volume = {245},
   number = {},
   pages = {175},
   year = {1958},
   doi= {}
}

@article{
 schneiderman:1969:pr:etf,
  title = {Velocity-Dependent Orbitals in Proton-On-Hydrogen-Atom Collisions},
  author = {Schneiderman, S. B. and Russek, A.},
  journal = PR,
  volume = {181},
  issue = {1},
  pages = {311--321},
  numpages = {0},
  year = {1969},
  month = {May},
  publisher = {American Physical Society},
  doi = {10.1103/PhysRev.181.311},
  url = {https://link.aps.org/doi/10.1103/PhysRev.181.311}
}

@article{
delos:1978:pra:etf,
  title = {Theory of near-adiabatic collisions. I. Electron translation factor method},
  author = {Thorson, W. R. and Delos, J. B.},
  journal = PRA,
  volume = {18},
  issue = {1},
  pages = {117--134},
  numpages = {0},
  year = {1978},
  month = {Jul},
  publisher = {American Physical Society},
  doi = {10.1103/PhysRevA.18.117},
  url = {https://link.aps.org/doi/10.1103/PhysRevA.18.117}
}

@article{
winter:1982:pra:etf,
  title = {Electron transfer in $p\ensuremath{-}{\mathrm{He}}^{+}$ and ${\mathrm{He}}^{2+}$-H collisions using a Sturmian basis},
  author = {Winter, Thomas G.},
  journal = PRA,
  volume = {25},
  issue = {2},
  pages = {697--712},
  numpages = {0},
  year = {1982},
  month = {Feb},
  publisher = {American Physical Society},
  doi = {10.1103/PhysRevA.25.697},
  url = {https://link.aps.org/doi/10.1103/PhysRevA.25.697}
}

@article{
errea:1994:etf,
doi = {10.1088/0953-4075/27/16/010},
url = {https://dx.doi.org/10.1088/0953-4075/27/16/010},
year = {1994},
month = {aug},
publisher = {},
volume = {27},
number = {16},
pages = {3603},
author = {L F Errea and  C Harel and  H Jouini and  L Mendez and  B Pons and  A Riera},
title = {Common translation factor method},
journal = {Journal of Physics B: Atomic, Molecular and Optical Physics}
}

@article{
  ohrn:1994:rmp:etf,
  title = {Time-dependent theoretical treatments of the dynamics of electrons and nuclei in molecular systems},
  author = {Deumens, E. and Diz, A. and Longo, R. and \"Ohrn, Y.},
  journal = RMP,
  volume = {66},
  issue = {3},
  pages = {917--983},
  numpages = {0},
  year = {1994},
  month = {Jul},
  publisher = {American Physical Society},
  doi = {10.1103/RevModPhys.66.917},
  url = {https://link.aps.org/doi/10.1103/RevModPhys.66.917}
}

@article{
  wu:2022:pssh,
  title={A phase-space semiclassical approach for modeling nonadiabatic nuclear dynamics with electronic spin},
  author={Wu, Yanze and Bian, Xuezhi and Rawlinson, Jonathan I and Littlejohn, Robert G and Subotnik, Joseph E},
  journal=JCP,
  volume={157},
  number={1},
  pages={011101},
  year={2022},
  publisher={AIP Publishing LLC}
}

@article{
  bian:2022:pssh,
  title={Modeling spin-dependent nonadiabatic dynamics with electronic degeneracy: a phase-space surface-hopping method},
  author={Bian, Xuezhi and Wu, Yanze and Rawlinson, Jonathan and Littlejohn, Robert G and Subotnik, Joseph E},
  journal={The Journal of Physical Chemistry Letters},
  volume={13},
  number={32},
  pages={7398--7404},
  year={2022},
  publisher={ACS Publications}
}

@article{
    wu:jcp:2021:first_attempt_complex,
    author = {Wu, Yanze and Subotnik, Joseph E.},
    title = "{Semiclassical description of nuclear dynamics moving through complex-valued single avoided crossings of two electronic states}",
    journal = JCP,
    volume = {154},
    number = {23},
    year = {2021},
    month = {06},
    issn = {0021-9606},
    doi = {10.1063/5.0054014},
    url = {https://doi.org/10.1063/5.0054014},
    note = {234101},
    eprint = {https://pubs.aip.org/aip/jcp/article-pdf/doi/10.1063/5.0054014/15977959/234101\_1\_online.pdf},
}

@article{
bian:jctc:2022:first_attempt,
author = {Bian, Xuezhi and Wu, Yanze and Teh, Hung-Hsuan and Subotnik, Joseph E.},
title = {Incorporating Berry Force Effects into the Fewest Switches Surface-Hopping Algorithm: Intersystem Crossing and the Case of Electronic Degeneracy},
journal = JCTC,
volume = {18},
number = {4},
pages = {2075-2090},
year = {2022},
doi = {10.1021/acs.jctc.1c01103},
    note ={PMID: 35263116},

URL = { 
        https://doi.org/10.1021/acs.jctc.1c01103
    
},
eprint = { 
        https://doi.org/10.1021/acs.jctc.1c01103
    
}

}

@article{
bian:2022:jcp:meaning,
    author = {Bian, Xuezhi and Qiu, Tian and Chen, Junhan and Subotnik, Joseph E.},
    title = "{On the meaning of Berry force for unrestricted systems treated with mean-field electronic structure}",
    journal = JCP,
    volume = {156},
    number = {23},
    pages = {234107},
    year = {2022},
    month = {06},
    issn = {0021-9606},
    doi = {10.1063/5.0093092},
    url = {https://doi.org/10.1063/5.0093092},
    eprint = {https://pubs.aip.org/aip/jcp/article-pdf/doi/10.1063/5.0093092/16670461/234107\_1\_online.pdf},
}

@article{
littlejohn:2022:jcp:parallel,
  title={The parallel-transported (quasi)-diabatic basis},
  author={Littlejohn, Robert and Rawlinson, Jonathan and Subotnik, Joseph},
  journal={The Journal of Chemical Physics},
  volume={157},
  pages={184303},
  year={2022},
  publisher={AIP Publishing}
}

@article{
littlejohn:2023:jcp:angmom,
  title={Representation and conservation of angular momentum in the Born--Oppenheimer theory of polyatomic molecules},
  author={Littlejohn, Robert and Rawlinson, Jonathan and Subotnik, Joseph},
  journal=JCP,
  volume={158},
  pages={104302},
  year={2023},
  publisher={AIP Publishing}
}

@article{
  krishna:2007:ehr_plus_berry,
  title={Path integral formulation for quantum nonadiabatic dynamics and the mixed quantum classical limit},
  author={Krishna, Vinod},
  journal={The Journal of chemical physics},
  volume={126},
  number={13},
  year={2007},
  publisher={AIP Publishing}
}

@article{
   takatsuka:2005:jcp,
   author = {Michiko Amano and Kazuo Takatsuka},
   title = {Quantum fluctuation of electronic wave-packet dynamics coupled with classical nuclear motions},
   journal = {The Journal of Chemical Physics},
   volume = {122},
   number = {8},
   ISSN = {0021-9606},
   DOI = {10.1063/1.1854115},
   url = {https://doi.org/10.1063/1.1854115},
   year = {2005},
   type = {Journal Article}
}

@article{
   athavale:2023:erf,
   author = {Vishikh Athavale and  Xuezhi Bian and  Zhen Tao and  Yanze Wu and  Tian Qiu and  Jonathan Rawlinson and  Robert G. Littlejohn and  Joseph E. Subotnik},
   title = {Surface Hopping, Electron Translation Factors, Electron Rotation Factors, Momentum Conservation, and Size Consistency},
   journal=JCP,
   volume = {159},
   number = {},
   pages = {114120},
   year = {2023},
   doi= {},
   note= {https://dx.doi.org/10.1063/5.0160965}
}

@article{
  nafie:1983:jcp:el_momentum,
  title={Adiabatic molecular properties beyond the Born--Oppenheimer approximation. Complete adiabatic wave functions and vibrationally induced electronic current density},
  author={Nafie, Laurence A},
  journal=JCP,
  volume={79},
  number={10},
  pages={4950--4957},
  year={1983},
  publisher={American Institute of Physics}
}

@article{
  gross:2022:prl:angmom,
  title={Energy, Momentum, and Angular Momentum Transfer between Electrons and Nuclei},
  author={Li, Chen and Requist, Ryan and Gross, E. K. U.},
  journal=PRL,
  volume={128},
  number={11},
  pages={113001},
  year={2022},
  publisher={APS}
}

@article{
   xuezhi:2023:total_ang_bomd,
   author = {Xuezhi Bian and Zhen Tao and Yanze Wu and Jonathan Rawlinson and Robert G. Littlejohn and Joseph E. Subotnik},
   title = {Total Angular Momentum Conservation in Ab Initio Born-Oppenheimer Molecular Dynamics},
   journal=PRB,
   volume = {108},
   number = {},
   pages = {L220304},
   year = {2023},  
   doi= {},
   note={https://dx.doi.org/10.1103/PhysRevB.108.L220304}
}

@article{
  littlejohn:2024:jcp:moyal,
    author = {Littlejohn, Robert and Rawlinson, Jonathan and Subotnik, Joseph},
    title = "{Diagonalizing the Born–Oppenheimer Hamiltonian via Moyal perturbation theory, nonadiabatic corrections, and translational degrees of freedom}",
    journal = JCP,
    volume = {160},
    number = {11},
    pages = {114103},
    year = {2024},
    month = {03},
    issn = {0021-9606},
    doi = {10.1063/5.0192465},
    url = {https://doi.org/10.1063/5.0192465},
    eprint = {https://pubs.aip.org/aip/jcp/article-pdf/doi/10.1063/5.0192465/19834737/114103\_1\_5.0192465.pdf},
}

@article{
  coraline:2024:jcp:ehrenfest_conserve,
  title={Total angular momentum conservation in Ehrenfest dynamics with a truncated basis of adiabatic states},
  author={Tao, Zhen and Bian, Xuezhi and Wu, Yanze and Rawlinson, Jonathan and Littlejohn, Robert G and Subotnik, Joseph E},
  journal=JCP,
  volume={160},
  number={5},
  year={2024},
  publisher={AIP Publishing}
}

@article{
  helgaker:2021:jcp:dynamics_in_magnetic_field,
  title={Ab initio molecular dynamics with screened Lorentz forces. I. Calculation and atomic charge interpretation of Berry curvature},
  author={Culpitt, Tanner and Peters, Laurens DM and Tellgren, Erik I and Helgaker, Trygve},
  journal=JCP,
  volume={155},
  number={2},
  year={2021},
  publisher={AIP Publishing}
}

@article{
  yanzewu:2024:jcp:pssh_conserve,
  title={Linear and angular momentum conservation in surface hopping methods},
  author={Wu, Yanze and Rawlinson, Jonathan and Littlejohn, Robert G and Subotnik, Joseph E},
  journal=JCP,
  volume={160},
  number={2},
  year={2024},
  page = { 024119 },
  publisher={AIP Publishing}
}

@article{
coraline:2024:jcp:pssh_conserve,
    author = {Tao, Zhen and Qiu, Tian and Bhati, Mansi and Bian, Xuezhi and Duston, Titouan and Rawlinson, Jonathan and Littlejohn, Robert G. and Subotnik, Joseph E.},
    title = {Practical phase-space electronic Hamiltonians for ab initio dynamics},
    journal = JCP,
    volume = {160},
    number = {12},
    pages = {124101},
    year = {2024},
    month = {03},
    issn = {0021-9606},
    doi = {10.1063/5.0192084},
    url = {https://doi.org/10.1063/5.0192084},
    eprint = {https://pubs.aip.org/aip/jcp/article-pdf/doi/10.1063/5.0192084/19846377/124101\_1\_5.0192084.pdf},
}

@misc{
      tian:2024:jcp:erf,
      title={A Simple One-Electron Expression for Electron Rotational Factors}, 
      author={Tian Qiu and Mansi Bhati and Zhen Tao and Xuezhi Bian and Jonathan Rawlinson and Robert G. Littlejohn and Joseph E. Subotnik},
      year={2024},
      eprint={2401.13778},
      archivePrefix={arXiv},
      primaryClass={physics.comp-ph},
      note={J. Chem. Phys., in press}
}

@article{
  helgaker:2022:jcp_conservation_laws_magnetic_field,
  title={Magnetic-translational sum rule and approximate models of the molecular Berry curvature},
  author={Peters, Laurens DM and Culpitt, Tanner and Tellgren, Erik I and Helgaker, Trygve},
  journal=JCP,
  volume={157},
  number={13},
  year={2022},
  publisher={AIP Publishing}
}

@article{
  cederbaum:1997:ijqc:magnetic_fied_thoughts,
  title={Molecules in strong magnetic fields: Some perspectives and general aspects},
  author={Schmelcher, P and Cederbaum, LS},
  journal=IJQC,
  volume={64},
  number={5},
  pages={501--511},
  year={1997},
  publisher={Wiley Online Library}
}

@article{
  gross:2017:prx:born_oppenheimer_mass,
  title = {On the Mass of Atoms in Molecules: Beyond the Born-Oppenheimer Approximation},
  author = {Scherrer, Arne and Agostini, Federica and Sebastiani, Daniel and Gross, E. K. U. and Vuilleumier, Rodolphe},
  journal = PRX, 
  volume = {7},
  issue = {3},
  pages = {031035},
  numpages = {23},
  year = {2017},
  month = {Aug},
  publisher = {American Physical Society},
  doi = {10.1103/PhysRevX.7.031035},
  url = {https://link.aps.org/doi/10.1103/PhysRevX.7.031035}
}

@misc{
    duston:2024:jctc_vcd,
      title={A Phase Space Approach to Vibrational Circular Dichroism}, 
      author={Titouan Duston and Zhen Tao and Xuezhi Bian and Mansi Bhati and Jonathan Rawlinson and Robert G. Littlejohn and Zheng Pei and Yihan Shao and Joseph E. Subotnik},
      year={2024},
      note = {https://arxiv.org/abs/2405.12404}
}

@article{
  waldeck:2024:chemrev:ciss,
  title={Chiral Induced Spin Selectivity},
  author={Bloom, Brian P and Paltiel, Yossi and Naaman, Ron and Waldeck, David H},
  journal=CR,
  year={2024},
  publisher={ACS Publications},
  volume = {124}, 
  pages = {1950-1991}
}

@article{
  case:2008:wigner_review,
  title={Wigner functions and Weyl transforms for pedestrians},
  author={Case, William B},
  journal={American Journal of Physics},
  volume={76},
  number={10},
  pages={937--946},
  year={2008},
  publisher={AIP Publishing}
}

@article{
    zhentao:2024:jcp:vcd_basis_free,
    author = {Tao, Zhen and Duston, Titouan and Pei, Zheng and Shao, Yihan and Rawlinson, Jonathan and Littlejohn, Robert and Subotnik, Joseph E.},
    title = {An electronic phase-space Hamiltonian approach for electronic current density and vibrational circular dichroism},
    journal = {The Journal of Chemical Physics},
    volume = {161},
    number = {20},
    pages = {204107},
    year = {2024},
    month = {11},
    issn = {0021-9606},
    doi = {10.1063/5.0233618},
    url = {https://doi.org/10.1063/5.0233618},
    eprint = {https://pubs.aip.org/aip/jcp/article-pdf/doi/10.1063/5.0233618/20267436/204107\_1\_5.0233618.pdf},
}

@article{
    bian:2024:pssh_translations_rotations,
    author = {Bian, Xuezhi and Wu, Yanze and Qiu, Tian and Tao, Zhen and Subotnik, Joseph E.},
    title = {A semiclassical non-adiabatic phase-space approach to molecular translations and rotations: Surface hopping with electronic inertial effects},
    journal = {The Journal of Chemical Physics},
    volume = {161},
    number = {23},
    pages = {234114},
    year = {2024},
    month = {12},
    issn = {0021-9606},
    doi = {10.1063/5.0242673},
    url = {https://doi.org/10.1063/5.0242673},
    eprint = {https://pubs.aip.org/aip/jcp/article-pdf/doi/10.1063/5.0242673/20312075/234114\_1\_5.0242673.pdf},
}

@misc{
 coraline:basisfree:2025,
 title = {A Basis-Free Phase Space Electronic Hamiltonian That Recovers Beyond Born-Oppenheimer Electronic Momentum and Current Density},
 author={Zhen Tao and  Tian Qiu and  Xuezhi Bian and  Joseph E. Subotnik},
 year={2025},
  journal=JCP,
  volume={162},
  pages={144111}
}

@article{
bian:2025:jctc:wigner_vibrations,
author = {Bian, Xuezhi and Khan, Cameron and Duston, Titouan and Rawlinson, Jonathan and Littlejohn, Robert G. and Subotnik, Joseph E.},
title = {A Phase-Space View of Vibrational Energies without the Born–Oppenheimer Framework},
journal = JCTC,
volume = {21},
number = {},
pages = {2880-2890},
year = {2025},
doi = {10.1021/acs.jctc.4c01294},
    note ={PMID: 40072941},

URL = { 
    
        https://doi.org/10.1021/acs.jctc.4c01294
    
    

},
eprint = { 
    
        https://doi.org/10.1021/acs.jctc.4c01294
    
    

}

}

@article{
  borgis:model:xuezhi:gross:wigner:cpl:2006,
  title={Generating approximate Wigner distributions using Gaussian phase packets propagation in imaginary time},
  author={Marinica, Dana Codruta and Gaigeot, Marie-Pierre and Borgis, Daniel},
  journal={Chemical physics letters},
  volume={423},
  number={4-6},
  pages={390--394},
  year={2006},
  publisher={Elsevier}
}

@book{
   tinkham_superconductivity,
   author = {M. Tinkham},
   title = {Introduction to Superconductivity},
   publisher = {Dover Publications},
   address = {},
   year = {2004}
}

@article{
littlejohn:flynn:1991:pra:coriolis,
  title = {Geometric phases in the asymptotic theory of coupled wave equations},
  author = {Littlejohn, Robert G. and Flynn, William G.},
  journal = {Phys. Rev. A},
  volume = {44},
  issue = {8},
  pages = {5239--5256},
  numpages = {0},
  year = {1991},
  month = {Oct},
  publisher = {American Physical Society},
  doi = {10.1103/PhysRevA.44.5239},
  url = {https://link.aps.org/doi/10.1103/PhysRevA.44.5239}
}

@article{bhati_magnetic_part2,
author = {Bhati, Mansi and Tao, Zhen and Bian, Xuezhi and Rawlinson, Jonathan and Littlejohn, Robert and Subotnik, Joseph E.},
title = {A Phase-Space Electronic Hamiltonian for Molecules in a Static Magnetic Field II: Quantum Chemistry Calculations with Gauge Invariant Atomic Orbitals},
journal = {The Journal of Physical Chemistry A},
volume = {129},
number = {20},
pages = {4573-4590},
year = {2025},
doi = {10.1021/acs.jpca.4c07905},
    note ={PMID: 40353803},

URL = { 
        https://doi.org/10.1021/acs.jpca.4c07905
},
eprint = { 
    
        https://doi.org/10.1021/acs.jpca.4c07905
}

}

@article{
bhati:2025:jpca:magnetic_part1,
author = {Bhati, Mansi and Tao, Zhen and Bian, Xuezhi and Rawlinson, Jonathan and Littlejohn, Robert and Subotnik, Joseph E.},
title = {A Phase-Space Electronic Hamiltonian for Molecules in a Static Magnetic Field. I: Conservation of Total Pseudomomentum and Angular Momentum},
journal = {The Journal of Physical Chemistry A},
volume = {129},
number = {20},
pages = {4555-4572},
year = {2025},
doi = {10.1021/acs.jpca.4c07904},
    note ={PMID: 40353811},

URL = {

        https://doi.org/10.1021/acs.jpca.4c07904



},
eprint = {

        https://doi.org/10.1021/acs.jpca.4c07904



}

}

@article{
xinchun:2025:wigner_one_state,
   author = {Xinchun Wu and  Xuezhi Bian and  Jonathan Rawlinson and  Robert G. Littlejohn and  Joseph E. Subotnik},
   title = {Recovering Exact Vibrational Energies Within a Phase Space Electronic Structure Framework},
   journal={},
   volume = {},
   number = {},
   pages = {},
   year = {},
   doi= {},
   note = {https://arxiv.org/abs/2506.08230}
}

@article{
xuezhi:cpr:review:2026,
    author = {Bian, Xuezhi and Duston, Titouan and Bradbury, Nadine and Tao, Zhen and Bhati, Mansi and Qiu, Tian and Wu, Xinchun and Wu, Yanze and Subotnik, Joseph E.},
    title = {The phase-space way to electronic structure theory and subsequently chemical dynamics},
    journal = {Chemical Physics Reviews},
    volume = {7},
    number = {1},
    pages = {011303},
    year = {2026},
    month = {01},
    issn = {2688-4070},
    doi = {10.1063/5.0286240},
    url = {https://doi.org/10.1063/5.0286240},
    eprint = {https://pubs.aip.org/aip/cpr/article-pdf/doi/10.1063/5.0286240/20867104/011303_1_5.0286240.pdf}
}

@book{
   bunker:book:2006,
   author = {Philp R. Bunker and Per Jensen},
   title = {Molecular Symmetry and Spectroscopy},
   publisher = {NRC Research Press},
   address = {},
   year = {2006}
}

@article{
polkovnikov:2026:pnas,
author = {Bernardo Barrera  and Daniel P. Arovas  and Anushya Chandran  and Anatoli Polkovnikov },
title = {The moving Born–Oppenheimer approximation},
journal = {Proceedings of the National Academy of Sciences},
volume = {123},
number = {7},
pages = {e2507816123},
year = {2026},
doi = {10.1073/pnas.2507816123},
URL = {https://www.pnas.org/doi/abs/10.1073/pnas.2507816123},
eprint = {https://www.pnas.org/doi/pdf/10.1073/pnas.2507816123}}

@article{
  born_oppenheimer_1927,
  title={Zur quantentheorie der molekeln},
  author={Born, Max and Oppenheimer, Robert},
  journal={Annalen der physik},
  volume={389},
  number={20},
  pages={457--484},
  year={1927},
  publisher={Wiley Online Library}
}

@article{
 moody:shapere:wilczek:1986,
  title = {Realizations of Magnetic-Monopole Gauge Fields: Diatoms and Spin Precession},
  author = {Moody, John and Shapere, A. and Wilczek, Frank},
  journal = PRL,
  volume = {56}, 
  issue = {9},
  pages = {893--896},
  numpages = {0},
  year = {1986},
  month = {Mar},
  publisher = {American Physical Society},
  doi = {10.1103/PhysRevLett.56.893},
  url = {https://link.aps.org/doi/10.1103/PhysRevLett.56.893}
}

@article{
  coraline:2026:roa,
  title={Non-resonant Raman optical activity from phase-space electronic structure theory},
  author={Tao, Zhen and Bhati, Mansi and Subotnik, Joseph E},
  journal={APL Computational Physics},
  volume={2},
  number={2},
  year={2026},
  publisher={AIP Publishing},
  page = {026101},
  doi = {10.1063/5.0315696},
  url = {https://doi.org/10.1063/5.0315696}
}

@article{
linqing:2026:lambda_doubling,
author = {Peng, Linqing and Qiu, Tian and Bradbury, Nadine and Bian, Xuezhi and Bhati, Mansi and Littlejohn, Robert and Kidwell, Nathanael M. and Subotnik, Joseph E.},
title = {Phase Space Electronic Structure Theory: From Diatomic Lambda-Doubling to Macroscopic Einstein–de Haas},
journal = JPCL,
volume = {17},
number = {10},
pages = {2799-2811},
year = {2026},
doi = {10.1021/acs.jpclett.5c03970},
    note ={PMID: 41771015},

URL = {
        
        https://doi.org/10.1021/acs.jpclett.5c03970



},
eprint = {
        
        https://doi.org/10.1021/acs.jpclett.5c03970



}
}

@article{
linqing:2026:spin_rotation,
    author = {Peng, Linqing and Duston, Titouan and Bradbury, Nadine and Bhati, Mansi and Tao, Xuecheng and Rosen, Michael B. and Subotnik, Joseph E.},
    title = {Conceptual Shift in
Our Understanding of Degenerate
Radical Spin Systems: Spin-Rotation Coupling Turned on Its Head},
    journal = {Journal of the American Chemical Society},
    volume = {148},
    number = {34},
    pages = {36583-36593},
    year = {2026},
    month = {08},
    issn = {0002-7863},
    doi = {10.1021/jacs.6c05615},
    url = {https://doi.org/10.1021/jacs.6c05615},
    eprint = {https://pubs.acs.org/jacsat/article-pdf/148/34/36583/66802912/jacs.6c05615.pdf}
}

@article{
  dou:2018:prb_vonOppen_noneq,
  title = {Universality of electronic friction. II. Equivalence of the quantum-classical Liouville equation approach with von Oppen's nonequilibrium Green's function methods out of equilibrium},
  author = {Dou, Wenjie and Subotnik, Joseph E.},
  journal = PRB,
  volume = {97}, 
  issue = {6},
  pages = {064303},
  numpages = {8},
  year = {2018},
  month = {Feb},
  publisher = {American Physical Society},
  doi = {10.1103/PhysRevB.97.064303},
  url = {https://link.aps.org/doi/10.1103/PhysRevB.97.064303}
}

@article{
polkovnikov:2017:quantum_geometry,
title = {Geometry and non-adiabatic response in quantum and classical systems},
journal = {Physics Reports},
volume = {697},
pages = {1-87},
year = {2017},
note = {Geometry and non-adiabatic response in quantum and classical systems},
issn = {0370-1573},
doi = {https://doi.org/10.1016/j.physrep.2017.07.001},
url = {https://www.sciencedirect.com/science/article/pii/S0370157317301989},
author = {Michael Kolodrubetz and Dries Sels and Pankaj Mehta and Anatoli Polkovnikov}
}

@article{
 mansi:nadine:moody_shapere_wilczek,
   author = {Mansi Bhati and  D. Vale Cofer-Shabica and  Jonathan I. Rawlinson and  Robert G. Littlejohn and  Joseph Subotnik and  Nadine C. Bradbury},
   title = {Electronic Structure in a Phase Space, non-Born-Oppenheimer Framework: Geometric Forces and Moody-Shapere-Wilzcek Revisited},
   journal={},
   volume = {},
   number = {},
   pages = {},
   year = {},
   doi= {},
   note={https://arxiv.org/abs/2605.27053}
}

@article{
vuilleumier:vcd:solidstate:2021,
author = {Jähnigen, Sascha and Zehnacker, Anne and Vuilleumier, Rodolphe},
title = {Computation of Solid-State Vibrational Circular Dichroism in the Periodic Gauge},
journal = JPCL,
volume = {12},
number = {30},
pages = {7213-7220},
year = {2021},
doi = {10.1021/acs.jpclett.1c01682},
    note ={PMID: 34310135},

URL = {
        
        https://doi.org/10.1021/acs.jpclett.1c01682



},
eprint = {
        
        https://doi.org/10.1021/acs.jpclett.1c01682



}

}

@article{
 satoh:roa:solid:2026:prl,
  title = {Raman Optical Activity Induced by Ferroaxial Order in ${\mathrm{NiTiO}}_{3}$},
  author = {Kusuno, Gakuto and Hayashida, Takeshi and Nagai, Takayuki and Watanabe, Hikaru and Oiwa, Rikuto and Kimura, Tsuyoshi and Satoh, Takuya},
  journal =  PRL,
  volume = {136},
  issue = {20},
  pages = {206902},
  numpages = {7},
  year = {2026},
  month = {May},
  publisher = {American Physical Society},
  doi = {10.1103/wrv8-4f7k},
  url = {https://link.aps.org/doi/10.1103/wrv8-4f7k}
}

@article{
  titouan:2025:ci_phase_space:jpcl,
  title={Conical intersections and electronic momentum as viewed from phase space electronic structure theory},
  author={Duston, Titouan and Bradbury, Nadine C and Tao, Zhen and Subotnik, Joseph E},
  journal=JPCL,
  volume={16},
  number={35},
  pages={8994--9003},
  year={2025},
  publisher={ACS Publications}
}

@article{
reiher:2016:static_dynamics_correlation,
author = {Stein, Christopher J. and von Burg, Vera and Reiher, Markus},
title = {The Delicate Balance of Static and Dynamic Electron Correlation},
journal = JCTC,
volume = {12},
number = {8},
pages = {3764-3773},
year = {2016},
doi = {10.1021/acs.jctc.6b00528},
    note ={PMID: 27409981},

URL = {
        
        https://doi.org/10.1021/acs.jctc.6b00528



},
eprint = {
        
        https://doi.org/10.1021/acs.jctc.6b00528



}
}

@book{
   born:huang,
   author = {Max Born and Kun Huang},
   title = {Dynamical Theory of Crystal Lattices},
   publisher = {Oxford University Press},
   address = {Oxford},
   year = {1954}
}

@article{
  becke:1988:jcp:beckegrid,
  title={A multicenter numerical integration scheme for polyatomic molecules},
  author={Becke, Axel D},
  journal=JCP,
  volume={88},
  number={4},
  pages={2547--2553},
  year={1988},
  publisher={American Institute of Physics}
}

@incollection{
  toddmartinez:book:beckegrid,
  title={dynamical quadrature grids: Applications in density functional calculations},
  author={Luehr, Nathan and Ufimtsev, Ivan and Martinez, Todd},
  booktitle={GPU Computing Gems Emerald Edition},
  pages={35--42},
  year={2011},
  publisher={Elsevier}
}

@article{
  frisch:cpl:1996:beckegrid,
  title={Achieving linear scaling in exchange-correlation density functional quadratures},
  author={Stratmann, R Eric and Scuseria, Gustavo E and Frisch, Michael J},
  journal=CPL,
  volume={257},
  number={3-4},
  pages={213--223},
  year={1996},
  publisher={Elsevier}
}

@article{
  stengel:traveling_pseudopoentials:prl:2026,
  title = {Rototranslational Sum Rules for Nuclear Dynamics via Traveling Pseudopotentials},
  author = {Stengel, Massimiliano and Royo, Miquel and Artacho, Emilio},
  journal = PRL,
  volume = {136},
  issue = {19},
  pages = {196401},
  numpages = {7},
  year = {2026},
  month = {May},
  publisher = {American Physical Society},
  doi = {10.1103/sdwb-qxrp},
  url = {https://link.aps.org/doi/10.1103/sdwb-qxrp}
}

@article{
mauri:2026:traveling_pseudopotentials,
  title={Non-adiabatic Ehrenfest dynamics with norm-conserving and ultra-soft pseudo-potentials with nuclear velocity corrections on the atomic orbitals within the Projector Augmented Wave Method framework},
  author={Fachin, Paolo and Macheda, Francesco and Barone, Paolo and Mauri, Francesco},
  journal={arXiv preprint arXiv:2606.06185},
  year={2026}
}

@article{
  teh:equilibrium:berryforce:prb,
  title={Antisymmetric Berry frictional force at equilibrium in the presence of spin-orbit coupling},
  author={Teh, Hung-Hsuan and Dou, Wenjie and Subotnik, Joseph E},
  journal=PRB,
  volume={104},
  number={20},
  pages={L201409},
  year={2021},
  publisher={APS}
}

@article{
  teh:non_equilibrium:berryforce:prb,
  title={Spin polarization through a molecular junction based on nuclear Berry curvature effects},
  author={Teh, Hung-Hsuan and Dou, Wenjie and Subotnik, Joseph E},
  journal=PRB,
  volume={106},
  number={18},
  pages={184302},
  year={2022},
  publisher={APS}
}

@article{
  heller:1976:wigner_problems,
  title={Wigner phase space method: Analysis for semiclassical applications},
  author={Heller, Eric J},
  journal=JCP,
  volume={65},
  number={4},
  pages={1289--1298},
  year={1976},
  publisher={American Institute of Physics}
}

@article{
    zain:shin_metiu_diabatization:2026_first_paper,
    author = {Zaidi, Zain and Bian, Xuezhi and Subotnik, Joseph E.},
    title = {Electron transfer, diabatic couplings, and vibronic energy gaps in a phase space electronic structure framework},
    journal = JCP,
    volume = {164},
    number = {19},
    pages = {194112},
    year = {2026},
    month = {05},
    issn = {0021-9606},
    doi = {10.1063/5.0325462},
    url = {https://doi.org/10.1063/5.0325462},
    eprint = {https://pubs.aip.org/aip/jcp/article-pdf/doi/10.1063/5.0325462/21008556/194112_1_5.0325462.pdf},
}

@article{
zain:inprogress,
   author = {Zain Zaidi and Yotam M. Feldman and Linqing Peng and  Joseph E. Subotnik},
   title = {An Exact Formulation of Phase-Space Electronic Structure Theory in One Dimension by Exploiting the Zero Curvature Condition},
   journal={},
   volume = {},
   number = {},
   pages = {},
   year = {2026},
   doi= {},
   note = "https://arxiv.org/html/2609.13120v1"}

@article{
curlee:becke:2026,
   author = {Ben Curlee and  Zheng Pei and  Xinchun Wu and Titouan Duston and  Todd J. Martinez and  Yihan
Shao and Joseph E. Subotnik},
   title = {Becke Weights as a Partitioning Scheme for Phase Space
Electronic Structure Theory},
   journal={},
   volume = {},
   number = {},
   pages = {},
   year = {2026},
   doi= {},
   note = {{\em submitted}}
}

@article{
hirshfeld:1977:weights,
  title={Bonded-atom fragments for describing molecular charge densities},
  author={Hirshfeld, Fred L},
  journal={Theoretica chimica acta},
  volume={44},
  number={2},
  pages={129--138},
  year={1977},
  publisher={Springer}
}

@article{
truhlar:mead:2003:ci_vs_avoided,
  title={Relative likelihood of encountering conical intersections and avoided intersections on the potential energy surfaces of polyatomic molecules},
  author={Truhlar, Donald G and Mead, C Alden},
  journal=PRA,
  volume={68},
  number={3},
  pages={032501},
  year={2003},
  publisher={APS}
}

\end{document}